\documentclass[acmsmall]{acmart}
\AtBeginDocument{%
  \providecommand\BibTeX{{%
    \normalfont B\kern-0.5em{\scshape i\kern-0.25em b}\kern-0.8em\TeX}}}

\setcopyright{acmcopyright}
\acmDOI{XXXXXXX.XXXXXXX}

\acmJournal{TKDD}

\usepackage{amsmath,amsfonts,amsthm}
\usepackage{caption}
\usepackage{subcaption}
\usepackage{mathtools}
\usepackage{algorithm, algorithmic}
\usepackage{graphicx}
\usepackage{textcomp}
\usepackage{xcolor}
\usepackage{enumitem}
\usepackage{nicefrac}
\usepackage{stfloats}
\usepackage[utf8]{inputenc}
\usepackage[english]{babel} 
\usepackage{multirow}
\usepackage{booktabs,tabularx} 
\newcommand\norm[1]{\left\lVert#1\right\rVert}
\usepackage{ulem}
\newtheorem{assumption}{Assumption}
\newtheorem{theorem}{Theorem}
\newtheorem{lemma}[theorem]{Lemma}

\newtheorem{remark}{Remark}
\newtheorem{definition}{Definition}
\newcommand{\tabincell}[2]{\begin{tabular}{@{}#1@{}}#2\end{tabular}}

\def\BibTeX{{\rm B\kern-.05em{\sc i\kern-.025em b}\kern-.08em
		T\kern-.1667em\lower.7ex\hbox{E}\kern-.125emX}}

\begin{document}

\title{LoSAC: An Efficient Local Stochastic Average Control Method for Federated Optimization}

%
%
%
%
%
%
%
\author{Huiming Chen}
\email{chenhuiming@mail.tsinghua.edu.cn}
\affiliation{%
	\institution{Department of Electronic Engineering, Tsinghua University}
	\city{Beijing}
	\country{China}
	\postcode{100084}
}

\author{Huandong Wang}
\affiliation{%
  \institution{Department of Electronic Engineering, Tsinghua University}
  \city{Beijing}
  \country{China}}
\email{wanghuandong@tsinghua.edu.cn}

\author{Quanming Yao}
\affiliation{%
	\institution{Department of Electronic Engineering, Tsinghua University}
	\city{Beijing}
	\country{China}}
\email{qyaoaa@tsinghua.edu.cn}

\author{Yong Li}
\authornotemark[1]
\affiliation{%
	\institution{Department of Electronic Engineering, Tsinghua University}
	\city{Beijing}
	\country{China}}
\email{liyong07@tsinghua.edu.cn}

\author{Depeng Jin}
\affiliation{%
	\institution{Department of Electronic Engineering, Tsinghua University}
	\city{Beijing}
	\country{China}}
\email{jindp@tsinghua.edu.cn}

\author{Qiang Yang}
\affiliation{%
	\institution{Department of Computer Science and Engineering, Hong Kong University of Science and Technology}
	\city{Hong Kong}
	\country{China}}
\email{qyang@cse.ust.hk}


\renewcommand{\shortauthors}{H. Chen, et al.}

\begin{abstract}
  Federated optimization (FedOpt), which targets at collaboratively training a learning model across a large number of distributed clients, 
  is vital for federated learning.
  {The primary concerns in FedOpt can be attributed to the model divergence and communication efficiency, which significantly affect the performance.}
  In this paper, we propose a new method,
  i.e., LoSAC, to learn from heterogeneous distributed data more efficiently.
  Its key algorithmic insight is to locally update the estimate for the global full gradient  after {each} regular local model update.
  Thus,
  LoSAC can keep clients' information refreshed in a more compact way. In particular, we have studied the convergence result for LoSAC.
  Besides, the bonus of LoSAC is the ability to defend the information leakage from the recent technique Deep Leakage Gradients (DLG). 
  Finally,
  experiments have verified the superiority of LoSAC
  comparing with state-of-the-art FedOpt algorithms. Specifically, LoSAC significantly improves communication efficiency by more than $100\%$ on average, mitigates the model divergence problem and equips with the defense ability against DLG.
\end{abstract}



\maketitle

\section{Introduction}
	Federated optimization (FedOpt) is essentially a distributed optimization in machine learning under the  specific setting that data is unevenly distributed over a large number of clients
\cite{ DBLP:journals/corr/KonecnyMRR16,DBLP:journals/corr/KonecnyMYRSB16,pmlr-v54-mcmahan17a,YangQiang2019FMLC,49232}. It has provided a feasible solution to collaboratively train a high-quality model without data exchanging. Thus, interests in applying FedOpt to  the major areas have been greatly increased, e.g., healthcare~\cite{Sheller,HUANG2019103291,BRISIMI201859} and smart city applications~\cite{wang2020federated,zheng2021federated}. 
Technically, the utmost goal of FedOpt is to reach a high communication efficiency  under the federated settings~\cite{DBLP:journals/corr/KonecnyMYRSB16,LiXiang2019OtCo,li2018federated,pmlr-v54-mcmahan17a}.

FedOpt  differs from the traditional distributed optimization~\cite{44634}  in data heterogeneity and client sampling respectively, i.e., the data is generally not independent and identical distribution (non-IID) due to each client's fashion,  and only partial clients perform multiple local updates before communication with the server in FedOpt. However, the recent studies have shown that {data heterogeneity (e.g., non-IID data)} will seriously degrade the performance in FedOpt~\cite{49232,li2018federated,Khaled,SCAFFOLD,ZhaoYue2018FLwN,LiXiang2019OtCo}. 
This can be attributed to the ``model divergence'' in FedOpt~\cite{SCAFFOLD,LiXiang2019OtCo}.
To be specific, multiple local update steps make each participating client approach to  the individual optimum targeting at optimizing the local loss function instead of the global one. Under the client sampling, this problem is worsened.
On one hand, several recent methods attempt to mitigate the model divergence problem~\cite{li2018federated,VRLSGD,SCAFFOLD}, however, there is still much room for the improvement in communication efficiency.
On the other hand, in FedOpt applications, many mobile devices may frequently go offline, and this  necessitates the demand for the fast convergent algorithm  to reach a complete learning~\cite{LiXiang2019OtCo}. 

To this end,  we summarize the two major challenges for the FedOpt design  \cite{Khaled,li2018federated,ZhaoYue2018FLwN,LiXiang2019OtCo}: 
(1). \textit{The model divergence problem}. It is mainly resulted from data heterogeneity and client sampling, and will lead to a slow and unstable convergence.
Since each client can only access to its own dataset, this makes the problem difficult to remedy; 
(2). \textit{The limited communication capability.} 
With a large number of clients, it requires the algorithm equipping with a fast speed. 
The recent representative methods for handling these challenges are MimeSVRG~\cite{MIME} and SCAFFOLD~\cite{SCAFFOLD}. 
They utilize the global variance reduction, which nevertheless may suffer from the outdated information.


LoSAC  incorporates the \textit{delayed gradients} {that estimate the global full gradient on clients.} Moreover,  they are updated with the newest information at each local iteration, and then aggregated with the information from the participated clients. The double estimates have improves the estimation accuracy, while  maintaining the low computation complexity. For further improving the communication efficiency, we adopt the common strategy in FedOpt to perform multiple local iterations.	While LoSAC has demonstrated its success to remedy the two key challenges in FedOpt, the performance improvements are prominent comparing to state-of-the-art FedOpt algorithms. 
The contributions of this work can be summarized as the following three aspects:
\begin{itemize}
	\item  A novel FedOpt algorithm is proposed, i.e., LoSAC. 
	We innovatively
	estimate the global full gradient on clients as the search direction, which targets at effectively handling the model divergence problem and substantially improving the communication efficiency. We further extend LoSAC to its proximal version and verifies the extraordinary performance. 
	
	\item We have studied the theoretical results of LoSAC, including the convergence result of $\mathcal{O}(\nicefrac{1}{RT})$, where $R$ and $T$ are communication round and local iteration respectively; the global variance reduction which guarantees the fast speed of LoSAC and the defense ability against DLG.
	
	\item 
	Extensive experiments are conducted in the settings of IID data and non-IID data. It has shown LoSAC equips with the strong capability to overcome the model divergence problem. Moreover,   it has exhibited quite high communication efficiency (i.e., more than $50\%$ fewer communication rounds than the state-of-the-art method to reach a specific accuracy), defense ability against DLG. In particular, its proximal version equips with the strong capability in solving the nonsmooth problem while the competitive algorithm fails to solve it.
	
\end{itemize}

The rest of the paper is organized as follows: Section II is the related work  discussing the existing popular FedOpt methods. Section III illustrates our proposed method by the motivation of the naive extension of SAGA, which only considers local gradient estimate. Then, we incorporate the global gradient estimate and formulate LoSAC. We further extend our proposed method to the proximal version for solving a wide class of nonsmooth problems. In Section IV, we study the theoretical properties of LoSAC, which has the global variance reduction on the search direction, is able to defense against DLG, and ensures the convergence and mitigates the model divergence problem. We conduct extensive experiments to verify the effectiveness of LoSAC in Section V.

\noindent
\textbf{Mathematical notations: } $[n]$ means the integer set $\{1:n\}$. $\Delta a:=a^+-a$ is presented as $a$'s increment when it has the updated value $a^+$. The gradient operator for a smooth function $f$ is denoted as $\nabla f$ and the statistical expectation is provided by $\mathbb{E}$. We use $l_2$-norm and for simplicity it is denoted as $\norm{\cdot}$. $\langle ,\rangle$ is the inner product. 
Moreover,  $f$ is called $L$-smoothness if $\norm{\nabla f(x)-\nabla f(y)}\leq L\norm{x-y}$,	where $L>0$ is the Lipschitz constant, and  that $f$ is strongly convex with $\mu>0$ satisfies $f(y)\geq f(x)+\left\langle \nabla f(x),y-x\right\rangle+\frac{\mu}{2}\norm{x-y}^2.$  $\text{prox}_{\beta f}(v)$ is the proximal operator defined in the following: $\text{prox}_{\beta f}(v)=\underset{x}{\text{armin } }\big( f(x)+\nicefrac{1}{2\beta}\norm{x-v}^2_2\big)$.
\section{Related Work}

\subsection{Stochastic Optimization}
\label{sec:co}

Considering there are $n$ data samples ($n$ is large), the stochastic optimization aims to solve
\begin{equation}
	\label{eq:sgd}
	\underset{x}{\text{min }}f(x) 
	= \frac{1}{n}\sum\nolimits_{i=1}^{n}f_i(x),
\end{equation}
where $x\in\mathbb{R}^d$ is the model, and $f_i$: $\mathbb{R}^d\rightarrow\mathbb{R}$ is the loss function with respect to the $i$th sample. 
One of the most popular method stochastic gradient descend (SGD)~\cite{RobbinsHerbert1951ASAM} utilizes a small mini-batch of samples $\mathcal{I}$ to calculate 
$g(x) = \nicefrac{1}{|\mathcal{I}|}\sum\nolimits_{i\in\mathcal{I}}f_i(x)$ 
and update the model $x$ via
\begin{align*}
	x^{t+1} = x^t-\eta\cdot g(x^t),
\end{align*}
where $\eta$ is the step-size.
Although $g(x)$ is an unbiased estimator of the full gradient $\nabla f(x)$, 
it may have large variance leading to a slow convergence~\cite{bottou2010large}. 
Thus,
how to control and reduce stochastic variance during mini-batch optimization is a central issue.

The variance reduction techniques~\cite{gower2020variance} have been developed to solve the above issue
and greatly accelerate the convergence of SGD.
Exemplar algorithms are
stochastic variance reduction gradient method (SVRG) \cite{NIPS2013_4937}, 
stochastic average gradient method (SAG) \cite{SchmidtMark2016Mfsw} and its extension SAGA~\cite{NIPS2014_5258}. 	
In particular,  
SAGA utilizes full gradient without 
{direct calculation}
and is instead updated with the newest partial information  at each iteration:
\begin{equation}\label{SAGA}
	x^{t+1}= x^t -\eta\cdot 
	\big\{
	\nabla f_j(x^t)-\nabla f_j(z^t_{j})+\frac{1}{n}\sum\nolimits_{i=1}^{n}\nabla f_{i}(z^t_{i})
	\big\},
\end{equation}
where $z_{j}$ is the \textit{delayed model} and is updated via:
\begin{equation}\label{z}
	z^{t+1}_{j}=\left\{
	\begin{aligned}
		&x^t_i, && \text{if $j$th data is sampled},\\
		&z^{t}_{j}, && \text{otherwise}.
	\end{aligned}
	\right.
\end{equation}
Let $\hat{\phi}^t := \sum\nolimits_{i=1}^{n}\nabla f_{i}(z^t_{i})$
and
$y^{t}_j=\nabla f_j(x^{t - 1})$.
Mimicking the efficient implementation in SAG~\cite{SchmidtMark2016Mfsw},  
and SAGA can be equivalently carried out in real applications as
\begin{equation}\label{loSAGA}
	x^{t+1}= x^t -\eta\cdot 
	\big\{
	\nabla f_j(x^t)-y^t_j+\frac{1}{n}\phi^t
	\big\},
\end{equation} 
where $\hat{\phi}$ is updated via 
\begin{equation}
	\hat{\phi}^{t+1} = \hat{\phi}^t - y^t_j + \nabla f_j(x^t).
\end{equation}
SAGA  has been shown the fast convergence speed. Moreover, it has the low computation level as SGD since only a single gradient is calculated during each update step.

Unfortunately,  
directly adapting these methods to FedOpt may not be effective since 
the fast convergence may cause the client 
quickly moving towards the individual optimum instead of the global one~\cite{LiXiang2019OtCo}.
\footnotesize
\begin{table*}[]
	
	\renewcommand{\arraystretch}{1.3}
	
	\caption{Summary of major federated optimization algorithms. PGD is the acronym of proximal gradient descent, divergence is abbreviated as div., PCP is the acronym of partial client participation,  comm. is short for communication.   Moroever, the convergence speed is provided when the iteration number and the participated clients number are fixed, $R$ is the communication round, $n_i$ is the local sample number and $d$ is the dimension. The details of the Defense against DLG (DaDLG) is in Section~\ref{sec:defense}.}
	\label{table_algm}
	\centering
	\begin{tabular}{l|l|l|l|l|l|l|l}
		\hline
		Algorithms&\tabincell{l}{Local update \\ step(s)} &\tabincell{l}{Local\\ Complexity}  &\tabincell{l}{Convergence\\speed}     &\tabincell{l}{Handling \\ model div.} &\tabincell{l}{Comm. \\efficiency} &\tabincell{l}{DaDLG} &PCP  \\ \hline
		
		FedAvg\cite{pmlr-v54-mcmahan17a} &\tabincell{l}{Multiple SGD \\ updates }  &$\mathcal{O}(Kd)$  &$\mathcal{O}(\nicefrac{1}{R})$     & $\times$ &Low &$\checkmark$ & $\checkmark$    \\ \hline
		SCAFFOLD\cite{SCAFFOLD} &\tabincell{l}{Multiple control \\ variate based\\ SGD updates}   &$\mathcal{O}(Kd)$  &$\mathcal{O}(\nicefrac{1}{R})$    &$\checkmark$     & High & $\checkmark$ &$\checkmark$   \\ \hline
		MimeSVRG \cite{MIME}  &\tabincell{l}{Multiple SVRG \\ updates }&$\mathcal{O}(Kd+n_id)$   &$\mathcal{O}(\nicefrac{1}{R})$   & $\checkmark$   & High & $\times$ &$\checkmark$   \\ \hline
		FedProx \cite{li2018federated} &\tabincell{l}{Single update \\ with PGD } &$\mathcal{O}(d)$  &$\mathcal{O}(\nicefrac{1}{R})$    &Limited & Medium & $\checkmark$  &$\times$ \\  \hline
		
		MFL~\cite{LiuWei2020AFLv} &\tabincell{l}{Momentum \\ gradient descent} & $\mathcal{O}(Kd)$ &$\mathcal{O}(\nicefrac{1}{R})$   & $\times$ & Low & $\times$ &$\times$  \\ \hline
		
		LoSAC (ours) &\tabincell{l}{Multiple global \\ SAGA based on\\ variance reduced \\ GD update} &\tabincell{l}{$\mathcal{O}(Kd)$} &$\mathcal{O}(\nicefrac{1}{R}) $  &$\checkmark$ &\tabincell{l}{Very \\ high}  & $\checkmark$ &$\checkmark$\\ \hline
	\end{tabular}
\end{table*}
\normalsize

\subsection{Federated Optimization (FedOpt)}

In~\cite{pmlr-v108-reisizadeh20a}, 
a quantized version of FedAvg, known as FedPaq, is proposed for reducing the message overload.  Mimicking the adaptation of SGD to FedOpt, the momentum gradient descent method which is a variant of SGD has been modified to fit for federated learning (MFL) \cite{LiuWei2020AFLv}. {Similarly, FedAdam~\cite{FedAdam} accommodated Adam~\cite{Adam} to FedOpt.}  However, these methods still suffer from the {model divergence} problem~\cite{WangShiqiang2019AFLi,ZhaoYue2018FLwN,LiXiang2019OtCo,DBLP:conf/aaai/YuYZ19}. A possible solution is to incorporate a quadratic restriction for the model divergence, which was known as FedProx~\cite{li2018federated}. Another solution may be the control variate for variance reduction, and based on which VRL-SGD has shown faster speed even with non-IID data~\cite{VRLSGD}. However, VRL-SGD does not support the client sampling which is more practical in FL. Furthermore, with the control variate, SCAFFOLD has shown the significant performance improvement for the data heterogeneity problem in FedOpt~\cite{SCAFFOLD}. Its core notion is to estimate the full gradient for the local search direction.
Most recently, a framework called Mime was proposed~\cite{MIME}, which adapts popular centralized algorithms (e.g., SGD, Adam etc.) to FedOpt.

As is mentioned in Section \ref{sec:co}, 
the naive adaptation of the variance reduction strategy in the federated settings may worsen the convergence. 
This is due to the reason that  variance reduction is applied for the local gradient estimation, 
which is biased from the global gradient.  
Both the SCAFFOLD and Mime framework are motivated by the idea of global variance reduction. 
However, the global full gradient estimates are kept over the whole local iterations, 
which may use outdated information
and thus these methods' capability to correct the model divergence is limited.
For LoSAC, 
both the local and global information are compactly utilized to keep the accurate estimate of the global gradient.
Thus LoSAC can reach the high quality of global variance reduction. 
For the details, 
we compare the recent FedOpt methods in Table~\ref{table_algm}.

	%
		%
		%
		%
		%
		%
		%

	\section{LoSAC Algorithm}
\label{sec:rel:fedopt}

In this section, we describe LoSAC to handle the major challenges in FedOpt. We first formulate the federated optimization problem, which aims to be solved by collaboration via many clients. Suppose there are $N$ clients, 
and each client $i\in[N]$ has the local loss function $f_i(x)$ with its own dataset $\mathcal{D}_i$ containing $n_i$ {samples}, i.e., $f_{i}(x)=\sum_{j=1}^{n_i}f_{i,j}(x)$, 
where $f_{i,j}(x)$ is a single loss function calculated by using the $j$th data in $\mathcal{D}_i$. 
Moreover, the total dataset over all clients are denoted by $\mathcal{D}$, 
i.e., $\mathcal{D}={\cup_{i\in[N]}}\mathcal{D}_i$.
As in the literature~\cite{pmlr-v54-mcmahan17a,SCAFFOLD,MIME,49232,DBLP:journals/corr/KonecnyMRR16},
the FedOpt aims to collaboratively solve the following empirical risk minimization problem over $N$ clients:
\begin{equation}\label{obj1}
	\underset{x}{\text{min }}f(x) 
	= \frac{1}{N}\sum\nolimits_{i=1}^{N}f_i(x),
\end{equation}
where  $f$ is the averaged loss function. The model as the optimization variable satisfies $x\in\mathbb{R}^d$. Moreover, the above functions satisfy $f:\mathbb{R}^d\rightarrow\mathbb{R}$, $f_i:\mathbb{R}^d\rightarrow\mathbb{R}$ and $f_{i,j}:\mathbb{R}^d\rightarrow\mathbb{R}$.  

\subsection{Naive Federated SAGA (FedSaga)}
\label{sec:naive}

For better illustration of LoSAC, we start with the simple extension of SAGA to the FedOpt (FedSaga).
Recall the local update step in FedAvg includes  multiple local SGD iterations. FedSaga simply reforms the local SGD  to the local SAGA. Specifically, given the local stochastic gradient $g_{i,j}$, 
FedSaga updates the local model $x_i$ at the $t$th local iteration via:
\begin{equation}\label{localSAGA}
	x^{t+1}_i= x^t_i -\eta\cdot 
	\big\{
	g_{i,j}(x^t_i)-g_{i,j}(z^t_{i,j})+\nabla f_i(\{z_i\})
	\big\},
\end{equation}
where $\nabla f_i(\{z_i\}):=\sum\nolimits_{j=1}^{n_i}\nabla f_{i,j}(z^t_{i,j})$ and we have denoted $\{z_i\}=\{z_{i,j}\}^{n_i}_{j=1}$, and $z_{i,j}$  is updated via:
\begin{equation}\label{fedsagaz}
	z^{t+1}_{i,j}=\left\{
	\begin{aligned}
		&x^t_i, && \text{if } j\text{th data in $\mathcal{D}_i$ is sampled},\\
		&z^{t}_{i,j}, && \text{otherwise}.
	\end{aligned}
	\right.
\end{equation}
A simple choice for $g_{i,j}$ is  $g_{i,j} = n_i \nabla f_{i,j}$. As shown in (\ref{fedsagaz}), $z_{i,j}$ is the delayed version of the local model if the $j$th data is not sampled, thus is called the \textit{delayed local model}.Moreover, it can be seen that the search direction in (\ref{localSAGA}) approaches the local gradient $\nabla f_i$ when the algorithm progresses to the optimum. 

Similarly to the SAGA update in (\ref{loSAGA}), we denote $\tilde{\phi}^t_i:=\sum\nolimits_{j=1}^{n_i}\nabla f_{i,j}(z^t_{i,j})$, which is the \textit{delayed local gradient}, and  store $\nabla f_{i,j}(z^{t}_{i,j})$ as $y^t_{i,j}$ on client $i$. 
Then, the local model can be equivalently updated in order via  (for simplicity we omit the superscript $t$):
\begin{equation}\label{fedsaga}
	(x_i,\tilde{\phi}_i,y_{i,j})\left\{
	\begin{aligned}
		&x_i\leftarrow x_i-\eta \{n_{i}\nabla f_{i,j}(x_i)-n_i y_{i,j}+\tilde{\phi}_i\},\\
		&\tilde{\phi}_i\leftarrow\tilde{\phi}_i-y_{i,j}+\nabla f_{i,j}(x^-_i),\\
		&y_{i,j}\leftarrow \nabla f_{i,j}(x^-_i),
	\end{aligned}
	\right.
\end{equation}
where $x^-_{i}$ is the local model before the local model update. 
Hence, we only need to calculate $\nabla f_{i,j}(x_i)$ for the local update, 
which makes FedSaga computationally efficient as SAGA. For the global update step, 
it aggregates the local models as FedAvg does, i.e., $x\leftarrow \nicefrac{1}{N}\sum x_i$.
We summarize FedSaga in Algorithm  \ref{alg:fedsaga}.

\begin{algorithm}[ht]
	\renewcommand{\algorithmicrequire}{\textbf{Input:}}
	\renewcommand{\algorithmicensure}{\textbf{Output:}}
	\caption{FedSaga} 
	\label{alg:fedsaga}
	\begin{algorithmic}[1]
		\STATE \textbf{server input:} The communication round $R$, initial $x$, $\phi$, $\Delta x_i=0$ for $i=[N]$.
		\STATE \textbf{client $i$'s input:} The local iteration number $T$, initial $\eta$, $x_i$, and $y_{i,j}$. 		
		\FOR{$r=1,\dots,R$}
		\STATE 
		
		\underline{\textbf{Server implements} steps 5-7:}
		
		\STATE
		Updates the $x$ and $\phi$ respectively:
		\STATE
		$x\leftarrow x+1/N\sum_{i\in{\mathcal{S}}}\Delta x_i$,
		
		\STATE Sample clients ${\mathcal{S}}\subseteq[N]$ and	transmit $x$ to client $i\in{\mathcal{S}}$.
		
		\STATE
		\underline{\textbf{Clients implement} steps 9-14 \textbf{in parallel for} $i\in{\mathcal{S}}$:}
		
		\STATE
		After receiving, set $x_i \leftarrow x$. 
		
		\FOR{$t=1,\dots,T$}
		
		\STATE
		Update $(x_i,\phi_{i}, y_{i,j})$ in order via (\ref{fedsaga}).
		
		\ENDFOR
		\STATE Calculate: $\Delta x_i\leftarrow x_i-x$.
		
		\STATE Client $i$ transmits $\Delta x_i$ to the server.
		
		\ENDFOR
	\end{algorithmic}
\end{algorithm}

\subsubsection{Limitation of FedSaga}
As shown in (\ref{localSAGA}), the variance reduction is realized by the local gradient $\nabla f_i(\{z_i\})$ on client $i$, which is biased from the global gradient $\nabla f(x)$. Moreover, the multiple local update steps will make the local model fast approach to the local individual optimum (based on the local loss function) instead of the global one~\cite{LiXiang2019OtCo} (See the empirical results that FedSaga even performs worse than FedAvg). 
Since only a small portion of the clients participate in the update on each communication round, 
the aggregated model will be further biased from the optimal global one. 
Intuitively, one can think of the global aggregation step in the aspect of the SGD update step, 
with the gradient only containing the partial information from the participated clients. 

\subsection{Local Stochastic Average Control}\label{sec:alg}

{As is discussed, FedSaga uses partial local information for the local update which results in bias from the global information. Hence, we propose LoSAC, which uses and updates the global information estimates  to make up for the bias.} To be specific, LoSAC updates the local model on each client $i\in\mathcal{S}$ via
\begin{equation}\label{LocalLoSAC}
	x_i
	\! \leftarrow \! 
	x_i 
	\! - \! 
	\eta\{\frac{1}{N} \sum_{n=1}^{N}\nabla f_n(\{z_n\})
	\! - \! 
	g_{i,j}(z_{i,j})
	\! + \!
	g_{i,j}(x_i)\},
\end{equation}
where $z_{i,j}$ is the \textit{delayed local model} and updated via (\ref{fedsagaz}). It can be seen the key difference between LoSAC and FedSaga is that LoSAC has used the estimate for the global information, i.e., $ \sum_{n=1}^{N}\nabla f_n(\{z_n\})$, while FedSaga only uses the local one.

\begin{figure}
	\centering
	\includegraphics[width=0.6\linewidth]{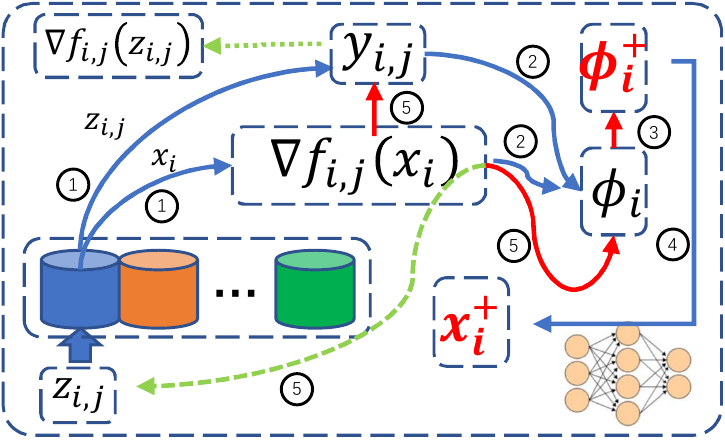}
	\caption{Illustration of local updates in LoSAC.  First, the local dataset block is randomly chosen for the gradient calculation $\nabla f_{i,j} (x_i )$. Second, note $y_{i,j}$ stores the delayed gradient $\nabla f_{i,j} (z_{i,j} )$, therefore, the delayed gradient $y_{i,j}$ can be replaced by the calculated gradient $ \nabla f_{i,j} (x_i )$. Third, with $\nabla f_{i,j} (x_i )$ and $y_{i,j}$,   the global gradient estimate $\phi_i^+$ is formulated. Fourth, the local update is performed to obtain $x_i^+$. Fifth, $y_{i,j}$ is updated with the gradient $\nabla f_{i,j} (x_i )$. }
	\label{fig:legend-losac}
\end{figure}

For the local step, we denote $\phi_i:=\sum_{n=1}^{N}\nabla f_n(\{z_n\})$, which is the \textit{delayed global gradient}. At the beginning of the local step, the server transmits $\phi$ to each client $i\in\mathcal{S}$ as the initialized $\phi_i$. Then, the following equations in order, which are equivalent to (\ref{LocalLoSAC}), 
are carried out multiple iterations:
\begin{equation}\label{losac_local}
	(x_i,\phi_i,y_{i,j})\left\{
	\begin{aligned}
		&x_i\leftarrow x_i - \eta
		\big\{ \nicefrac{1}{N} \phi_i -y_{i,j}+g_{i,j}(x_i) \big\},\\
		&\phi_i\leftarrow\phi_i-y_{i,j}+g_{i,j}(x^-_{i}),\\
		&y_{i,j}\leftarrow g_{i,j}(x^-_i).
	\end{aligned}
	\right.
\end{equation}
After the local update step, except sending the updated quantity $\Delta x_i$ to the server as FedSaga, 
client $i$ also needs to send $\Delta\phi_i:=\phi^+_i-\phi$ for further aggregation. 
Here, $\phi^+_i$ is the updated $\phi_i$ after the local update step. Moreover, although $\phi_i$ contains all the \textit{delayed local gradients} $\nabla f_{[1:N]}$, 
client $i$ can only update its own $\nabla f_{i}$ while others are remained unchanged, 
owing to the FedOpt setting that client $i$ can only have access to its own dataset $\mathcal{D}_i$,  i.e.,
\begin{equation}\label{globalagg}
	\Delta\phi_i=\phi^+_i-\phi_i=\nabla f_i(\{z^+_i\})-\nabla f_i(\{z_i\}).
\end{equation}
The local update procedure can be illustrated in Fig. \ref{fig:legend-losac}. 

For the global step, the server receives all the update quantities $(\Delta x_i, \Delta \phi_i)$ for $i\in\mathcal{S}$ and performs the aggregation:
\begin{equation}\label{globalupdate}
	\begin{aligned}
		x&\leftarrow x 
		+ \nicefrac{1}{N}
		\sum_{i\in{\mathcal{S}}}\Delta x_i, \text{ and }  \phi\leftarrow \phi
		+ \nicefrac{N}{S}
		\sum_{i\in{\mathcal{S}}}\Delta \phi_i.
	\end{aligned}
\end{equation}

While $\phi_i$ is compactly updated with the local information, it is also aggregated with the global information. Therefore, the twice estimates make $\phi_i$  reach an accurate estimate for the global gradient. The detail of LoSAC is summarized in Algorithm \ref{alg:losac}. 
\begin{algorithm}[ht]
	\renewcommand{\algorithmicrequire}{\textbf{Input:}}
	\renewcommand{\algorithmicensure}{\textbf{Output:}}
	\caption{LoSAC} 
	\label{alg:losac}
	\begin{algorithmic}[1]
		\STATE \textbf{server input:}  initial $x$, $\phi$, $\Delta x_i=0$ and $\Delta \phi_i=0$ for $i=[N]$.
		\STATE \textbf{client $i$'s input:} initial $\eta$ and $y_{i,j}$.
		
		\FOR{$r=1,\dots,R$}
		\STATE 
		
		\underline{\textbf{Server implements} steps 5-6:}
		
		\STATE
		Update $(x,\phi)$ via (\ref{globalupdate}).

		\STATE 
		
		Sample clients ${\mathcal{S}}\subseteq[N]$ and	transmit $(x,\phi)$ to client $i\in{\mathcal{S}}$.
		
		\STATE
		\underline{\textbf{Clients implement} steps 8-14 \textbf{in parallel for} $i\in{\mathcal{S}}$:}
		
		\STATE
		After receiving, set $x_i \leftarrow x$ and $\phi_i \leftarrow \phi$. 
		
		\FOR{$t=1,\dots,T$}
		
		\STATE Sample index $j$ from $[n_i]$.
		\STATE
		Update $(x_i, \phi_i, y_{i,j})$ in order via (\ref{losac_local}).
		
		\ENDFOR
		\STATE Calculate: $\Delta x_i\leftarrow x_i-x$ and $\Delta \phi_i\leftarrow \phi_i-\phi$.
		
		\STATE Client $i$ transmits $(\Delta x_i, \Delta \phi_i)$ to the server.

		\ENDFOR
	\end{algorithmic}
\end{algorithm}

\begin{remark}
	In real applications, a block of data points can be bundled for evaluating a single loss function $f_{i,j}$ and in this way, 
	the memory cost on client $i$ will be  $O(\lceil \nicefrac{n_id}{b} \rceil)$, 
	where $b$ is the number of data points in each block.
	
\end{remark}


\subsection{Extension with Proximal Operator}
Proximal operator has been shown as an effective tool for solving  nonsmooth, constrained, large-scale, or distributed
problems \cite{DBLP:journals/ftopt/ParikhB14}. In this subsection, we extend our proposed method with proximal operator for solving a wider class of problems, such as $l_1$ regularization or low-rank matrix estimation. Specifically, the problem under federated settings can be formulated in the following:
\begin{equation}
	\underset{x\in\mathbb{R}^d}{\text{min  }} \frac{1}{N}\sum_{i=1}^{N}f_i(x)+\Psi(x), 
\end{equation}
where $x\in\mathbb{R}^d$ is the model, $f_i(x):\mathbb{R}^d\rightarrow\mathbb{R}$ is the local loss function, and $\Psi:\mathbb{R}^d\rightarrow \mathbb{R}$ is the nonsmooth convex regularizer. When  $f_i(x) = \nicefrac{1}{n_i}\sum_{(d_i,y_i)\in\mathcal{D}_i}\norm{x^Td_i-y_i}^2$, where $\mathcal{D}_i$ is the local dataset on client $i$, and $\Psi=\lambda\norm{\cdot}_1$, the problem is known as LASSO \cite{lasso}. As for the local gradient descent based proximal step, it can be derived:
\begin{equation}\label{pgd}
	x^+_i\leftarrow \text{argmin}_{z} \frac{1}{2\eta}\norm{z-(x_i-\eta\nabla f_i(x_i))}+\Psi(x_i),
\end{equation}
which can be solved via $x^+_i\leftarrow \text{prox}_{\eta\Psi}(x_i-\eta\nabla f_i(x_i))$. Mimicking the PGD step, we incorporate the global gradient estimate in LoSAC to replace the local gradient in (\ref{pgd}), which subsequently lead to 
\begin{equation}\label{losac-prox}
	x_i\leftarrow \text{prox}_{\eta\Psi}(x_i - \eta
	\big\{ \nicefrac{1}{N} \phi_i -y_{i,j}+g_{i,j}(x_i) \big\}). 
\end{equation}	
While adopting  (\ref{losac-prox}), we maintain the local estimate for  $\phi_i$ and the global aggregation for $(x,\phi)$, the resulting algorithm is called LoSAC-Prox. In Section \ref{sec:ne}, we will apply LoSAC-Prox for solving the low-rank matrix estimation problem for further showing the superiority over the state-of-the-art algorithm.


\section{Theoretical Analysis}

In this section,  
the theoretical analysis of LoSAC is presented. 
Specifically, we first provide the global variance reduction of LoSAC to show that the variance of the search direction is vanishing. Moreover, we also analyze the enhanced defense ability of LoSAC in gradient leakage. Then we study the convergence analysis, of which 
one challenge is resulted from  the multiple local iterations. Moreover, it can be seen in (\ref{LocalLoSAC}) that the \textit{delayed local models} $\{z_{i}\}$ of client $i$ make the convergence analysis  difficult for evaluation, since after a few iterations, the \textit{delayed local gradient} $\nabla f_i(\{z_i\})$ has the mixed arguments. Finally, we show LoSAC equips with the  ability in handling model divergence.

	%
	%
	%
	%
	%

\subsection{Global Variance Reduction}
\label{ssec:vard}

The variance of the search direction in LoSAC is progressively reduced to zero comparing to the {SGD} update in FedAvg. 
Moreover, it maintains the robustness in LoSAC for the convergence. 
Comparing with the recent method MimeSVRG \cite{MIME} and SCAFFOLD \cite{SCAFFOLD}, 
LoSAC equips with the benefit of compactly refreshed global variance reduction. 
Hence its convergence performance can be improved comparing to MimeSVRG and SCAFFOLD, 
which is demonstrated in the numerical experiments.  
In the following, we provide the global variance reduction in Lemma \ref{lemma6}:

\begin{lemma}\label{lemma6}
	Suppose the sequence $\{x^r\}$ generated by Algorithm 1 is expected to converge, i.e., $\mathbb{E}\norm{x^r-x^*}\rightarrow 0$. Moreover, for the $i$th client, denote the search direction in the local update as $\tilde{g}^t_i=\nicefrac{1}{N} \phi^t_i -g_{i,j_t}(z^t_{i,j_t})+g_{i,j_t}(x^t_i)$. Then, the variance of the search direction $\tilde{g}^t_i$ is progressively vanished, i.e., $\mathbb{E}\|{\tilde{g}^t_i-\mathbb{E}(\tilde{g}^t_i)}\|^2\rightarrow 0$.
\end{lemma}


\subsection{Defense Ability to the Gradient Leakage} 
\label{sec:defense}

An important benefit of  LoSAC is the enhancement of
the defense ability to  the recent technique 
\textit{Deep Leakage from Gradients (DLG)}~\cite{NIPS2019_9617}
which aims to obtain the information leakage from the gradient. 
As the illustration,
we denote 
$f_i(x;\mathcal{D}_i)=f_i(x)$ 
to explicitly emphasize the dependency on the input sample $\mathcal{D}_i$. 
Then, the DLG is defined as follows: 

\begin{definition}[DLG~\cite{NIPS2019_9617}]
	For an algorithm $\mathcal{A}$, let the associated gradient be $\nabla f_i(x;\mathcal{D}_i)$ and
	the model parameter be $x$,  
	and 
	\begin{equation*}
		\mathcal{D}'_i
		= 
		\arg\min\nolimits_{\tilde{\mathcal{D}}_i}
		\norm{\nabla f_i(x; \tilde{\mathcal{D}}_i)-\nabla f_i(x; \mathcal{D}_i)}^2.
	\end{equation*}
	If a Malicious Attacker (MA)  is able to obtain $\mathcal{D}_i$  
	by finding $\mathcal{D}'_i$ above,
	then the algorithm $\mathcal{A}$ suffers from Deep Leakage from Gradients (DLG).
	%
\end{definition}

Hence, with the technique of DLG, 
MA is able to progressively match the gradient $\nabla f_i(x,\mathcal{D}_i)$ by  minimizing the difference between the ``dummy gradient'' $\nabla f_i(x, \tilde{\mathcal{D}}_i)$ and  $\nabla f_i(x,\mathcal{D}_i)$. 
Moreover,  when the optimum is reached, MA steals the data $\mathcal{D}_i$ from the $i$th client, i.e., 
$\mathcal{D}'_i=\mathcal{D}_i$. 

According to the above definition, MA is not able to apply DLG algorithm to obtain the local dataset in LoSAC. We illustrate this from two reasons. First, the $i$th client's gradient is evaluated at many \textit{delayed local models}  $\{z_i\}$,  i.e., $\nabla f_i(\{z^t_i\},\mathcal{D}_i)$. Moreover, $\{z_i\}$ are stored locally that MA is difficult to obtain; second, LoSAC transmits the increments $\Delta x_i$  and $\Delta \phi_i$  instead of the gradients. Howover, the distributed SGD~\cite{NIPS2019_9617} and Mime framework~\cite{MIME} communicates the local gradient with the the local model, i.e., $x$ and $\nabla f_i(x;\mathcal{D}_i)$, hence MA is able to steal the local data $\mathcal{D}_i$ by DLG technique.

\subsection{Convergence Result}
In this subsection, we study the convergence property of our proposed method. We first show the progress of each communication round in  Lemma \ref{lemma4}. Particularly, we need the following regular assumptions. 
\begin{assumption}\label{ass:f}
	Following assumptions are made:
	\begin{itemize}
		\item [A1.] The gradient estimates have bounded variance, i.e., $\text{Var}[g(x)]\leq\sigma_f$ and $\text{Var}[g_{i,j}(x)]\leq\sigma_f$,		
		where $g(x)$ and $g_{i,j}(x)$ are the unbiased estimates of $\nabla f(x)$ and $\nabla f_i(x)$ respectively.
		
		\item [A2.]
		The second-order moments of all the unbiased gradient estimates are bounded, i.e.,
		$\mathbb{E}\norm{g(x)}^2$
		$\leq\delta_f$ and $\mathbb{E}_j\norm{g_{i,j}(x)}^2\leq\delta_f$.
	\end{itemize}
\end{assumption}

Here,
Assumptions A1 and A2 have been regularly made for convergent analysis in optimization literatures~\cite{Khaled,DBLP:conf/iclr/Stich19,NIPS2018_7697,JMLR:v14:zhang13b,LiXiang2019OtCo} and~\cite{NIPS2018_7697,LiXiang2019OtCo,NIPS2018_7837,JMLR:v14:zhang13b}, respectively. Moreover,  Note that Assumptions A1 and A2 imply that the second-order moments of the gradients $\nabla f(x)$ and $\nabla f_i(x)$ are also bounded, i.e., $\mathbb{E}\norm{\nabla f(x)}^2\leq\delta_f-\sigma_f$ and $\mathbb{E}\norm{\nabla f_i(x)}^2\leq\delta_f-\sigma_f$. 

From (\ref{LocalLoSAC}), we intuitively have the approximate gradient descent (GD) step in each local iteration 
\begin{equation}
	x^{t+1}_i\simeq x^{t}_i - \frac{\eta}{N} \sum\nolimits_{n=1}^{N}\nabla f_n(\{z^t_n\}).
\end{equation}
Hence, the local iteration progress can be bounded above with reference to the GD theory \cite{naivesaga} and subsequently, the one round progress  can be obtained in the following:

\begin{lemma}\label{lemma4}
	Suppose functions $f$, $\{f_i\}$ and $\{f_{i,j}\}$ are strongly convex and $L$-smooth that satisfy Assumption~\ref{ass:f}, 
	and denote $x^*$ as the optimal point, 
	if $T$ is sufficiently large \footnote{We assume $T$ is sufficiently large such that the locally stored $\{z_{i,j}\}$ is updated at least once.}, 
	there exist positive variables $h_2$, $\lambda'_2$ and $\nu'_2$ such that
	\begin{align}
		\mathbb{E}\|x^{r+1}-x^*\|^2&\leq\frac{-2\eta ST}{N}\mathbb{E}\{f(x^r)-f(x^*)\}+(1-\eta h_2)\mathbb{E}\|x^{r}-x^*\|^2+\lambda'_2\eta^2+\nu'_2\eta^3. \label{bound2}
	\end{align}
\end{lemma}

It should be noted that  the assumption that $f_{i,j}$ is strongly convex is strong in real applications, e.g., $f_{i,j}$ is the loss function in a neural network, but   this assumption  can be simply realized by appending a $l_2$ regularization term to $f_{i,j}$ to form the strongly convex function. 

Based on Lemma \ref{lemma4}, the convergence speed can be obtained in the following.

\begin{theorem}\label{theorem1}
	Suppose functions $f$, $\{f_i\}$ and $\{f_{i,j}\}$ satisfy \textit{A1} and \textit{A2}. 
	Given a positive sequence $\{w_r\}$ defined by $w_r=(1-\eta h_2)^{-r}$,  LoSAC has a weighted averaging convergence speed:
	\begin{equation}
		\frac{1}{W_R}\sum\nolimits_{r=0}^{R}w_r\mathbb{E}\{f(x^r)-f(x^*)\}\leq\mathcal{O}(\frac{N\lambda'}{2SThR} ).
	\end{equation}
	where $W_R:=\sum\nolimits_{r=0}^{R}w_r$, $h$ and $\lambda'$ are positive. 
\end{theorem}


It can be seen the convergence speed will be faster if there are more local iterations $T$	while the computation complexity is also increased. 
Furthermore, if more clients (larger $|\mathcal{S}|$)  are participated in model training, it will be faster for convergence. 	Moreover,  our analysis does not assume data heterogeneity while~\cite{SCAFFOLD} does. 
This is due to the reason that $\phi_i$ estimates the global information. 
We intuitively and empirically illustrate this in Section \ref{modeldiv} and numerical experiments respectively.

\subsection{Handling Model Divergence}
\label{modeldiv}

LoSAC is expected to equip with the capability to overcome the model divergence problem that resulted from data heterogeneity and  client sampling in FedOpt. 
As an intuitive illustration of this, the local update step on client $i$ in expectation can be approximated as $\mathbb{E}_t(x^{t+1}_i)\simeq x^{t}_i - \frac{\eta}{N} \sum\nolimits_{n=1}^{N}\nabla f_n(\{z^t_n\})$, 	
where we have assumed $g_{i,j_t}(z^t_{i,j_t})\simeq g_{i,j_t}(x^t_{i})$. 
It mimics the full gradient descent step. To a certain extent, 
appending the term $\nicefrac{1}{N} \phi^t_i-y_{i,j_t}$ in the local update step makes up the deviation for $g_{i,j_t}(x^t_i)$ from the full gradient. 
Thus, the model divergence problem can be relatively mitigated.

\section{Numerical Experiments}\label{sec:ne}


\subsection{Experimental Settings} 

\subsubsection{Training model} 
We choose 2NN in~\cite{pmlr-v54-mcmahan17a}, logistic regression and  low rank matrix estimation \cite{lrme}
as the training models. Specifically, 2NN is a fully connected neural network with 2 hidden layers with $200$ ReLU activation functions in the each hidden layer and a softmax output. 

\subsubsection{Datasets} 
Three real datasets are chosen for overall performance, ablation study and DLG study, namely MNIST~\cite{726791}, Human Activity Recognition Using Smartphones dataset (HAR) \cite{HAR} and Epileptic Seizure Recognition dataset (ESR) \cite{ESR}. We choose MNIST dataset since it has been widely applied for the study. Moroever, HAR and ESR are chosen due to the increasing interests and the large potential for the FedOpt applications in mobile devices and healthcare, respectively. Specifically,  we use $60,000$ for training and $10,000$ for testing in MNIST, $7,352$ for training and $2,947$ for testing in HAR, and   $9,200$ for training and $2,300$ for testing in ESR. Moreover, we choose synthetic dataset for the low rank matrix estimation.  

\scriptsize
\begin{table*}[]
	\renewcommand{\arraystretch}{1.3}
	
	\caption{Ablation study: measured by the communication rounds for LoSAC and  SCAFFOLD to reach a specific accuracy (85\% for all datasets), loss and test accuracy. We consider the different local iteration $T$ and the local dataset division $M$ ($M$ also corresponds to the local memory size for storing $y_{i,j}$) for calculating the gradient for comparisons to show the efficient computation and communication of LoSAC. Other parameters are set as $S=10$ and $\eta=10^{-4}$. Moreover, the accuracy is measured at round $500$, which is sufficient for reaching a satisfactory accuracy.} 
	\label{table_datasets}
	\centering
	\begin{tabular}{l|l|l|l|l|l|l|l|l|l|l}
		\hline
		Datasets&Methods  &\tabincell{l}{Rounds\\ ($T=2$)}  &\tabincell{l}{Loss\\ ($T=2$)}  &\tabincell{l}{ACC\\ ($T=2$)}  &\tabincell{l}{Rounds\\ ($T=4$)}   &\tabincell{l}{Loss\\ ($T=4$)}   &\tabincell{l}{ACC\\ ($T=4$)}   &\tabincell{l}{Rounds\\ ($T=6$)}   &\tabincell{l}{Loss\\ ($T=6$)}   &\tabincell{l}{ACC\\ ($T=6$)}   \\ \hline
		\multirow{6}{*}{MNIST} &\tabincell{l}{SCAFFOLD \\ ($M=2$) } &64$(1\times)$  &0.1853  &  $94.37\%$& 47$(1\times)$  & 0.1268 & $96.00\%$  &39$(1\times)$  & 0.1051 &$96.41\%$  \\ \cline{2-11} 
		&\tabincell{l}{SCAFFOLD \\ ($M=3$) }  &63$(1\times)$  & 0.1958 &$94.08\%$  &39$(1\times)$  & 0.1392 &$95.44\%$  &37$(1\times)$  & 0.1109 &$95.94\%$  \\ \cline{2-11} 
		&\tabincell{l}{SCAFFOLD \\ ($M=5$) }  &57$(1\times)$  &0.1847  &$94.38\%$  &44$(1\times)$  &0.1253  &$96.03\%$  &43$(1\times)$  &0.1061  &$96.40\%$ \\ \cline{2-11} 
		&\tabincell{l}{LoSAC \\ ($M=2$) } &56$\boldsymbol{(1.14\times)}$  &$\boldsymbol{0.1310}$  &$\boldsymbol{95.91}\%$  &38$\boldsymbol{(1.24\times)}$  &$\boldsymbol{0.0840}$ &$\boldsymbol{96.83}\%$  &30$\boldsymbol{(1.30\times)}$  &$\boldsymbol{0.0703}$  &$\boldsymbol{96.97}\%$  \\ \cline{2-11} 
		&\tabincell{l}{LoSAC \\ ($M=3$) }   &46$\boldsymbol{(1.37\times)}$  &$\boldsymbol{0.1162}$  &$\boldsymbol{95.96}\%$  &32$\boldsymbol{(1.22\times)}$  &$\boldsymbol{0.0632}$  &$\boldsymbol{96.99}\%$  & 34$\boldsymbol{(1.09\times)}$ &$\boldsymbol{0.0438}$  &$\boldsymbol{97.36}\%$  \\ \cline{2-11} 
		&\tabincell{l}{LoSAC \\ ($M=5$) }  &44$\boldsymbol{(1.30\times)}$  &$\boldsymbol{0.0887}$   &$\boldsymbol{96.45}\%$  & 34$\boldsymbol{(1.29\times)}$ &$\boldsymbol{0.0501}$  &$\boldsymbol{97.26}\%$  &26$\boldsymbol{(1.65\times)}$  &$\boldsymbol{0.0339}$  &$\boldsymbol{97.53}\%$  \\ \hline
		\multirow{6}{*}{HAR} &\tabincell{l}{SCAFFOLD \\ ($M=2$) } &177$(1\times)$  &$0.2244$  &$90.19\%$  &175$(1\times)$  &$0.1558$  &$92.87\%$  &170$(1\times)$  &$0.1479$  &$92.53\%$  \\ \cline{2-11} 
		&\tabincell{l}{SCAFFOLD \\ ($M=3$) }  &158$(1\times)$  &$0.2093$  &$91.45\%$  &161$(1\times)$  &$0.1421$  &$92.74\%$  &166$(1\times)$  &$\boldsymbol{}$0.1334  & $92.40\%$  \\ \cline{2-11} 
		&\tabincell{l}{SCAFFOLD \\ ($M=5$) }  &137$(1\times)$  &$\boldsymbol{}$0.2178  &$91.08\%$  &156$(1\times)$  &$\boldsymbol{}$0.1518  &$93.65\%$  &122$(1\times)$  &$\boldsymbol{}$0.1373  &$91.92\%$   \\ \cline{2-11} 
		&\tabincell{l}{LoSAC \\ ($M=2$) } &102$\boldsymbol{(1.74\times)}$  &$\boldsymbol{0.1611}$  & $\boldsymbol{92.06}\%$ &47$\boldsymbol{(3.72\times)}$  &$\boldsymbol{0.1253}$  &$\boldsymbol{94.10}\%$  &61$\boldsymbol{(2.79\times)}$  &$\boldsymbol{0.1176}$  & $\boldsymbol{93.76}\%$ \\ \cline{2-11} 
		&\tabincell{l}{LoSAC \\ ($M=3$) }   &74$\boldsymbol{(2.14\times)}$  &$\boldsymbol{0.1185}$  &$\boldsymbol{93.52}\%$  &44$\boldsymbol{(3.66\times)}$  &$\boldsymbol{0.1014}$  &$\boldsymbol{93.82}\%$  &35$\boldsymbol{(4.74\times)}$  &$\boldsymbol{0.1035}$  & $\boldsymbol{92.77}\%$ \\ \cline{2-11} 
		&\tabincell{l}{LoSAC \\ ($M=5$) }  &89$\boldsymbol{(1.54\times)}$  &$\boldsymbol{0.1076}$  & $\boldsymbol{93.52}\%$ &55$\boldsymbol{(2.84\times)}$  &$\boldsymbol{0.0902}$  &$\boldsymbol{94.77}\%$  &43$\boldsymbol{(2.84\times)}$  &$\boldsymbol{0.0921}$  & $\boldsymbol{94.33}\%$ \\ \hline
		\multirow{6}{*}{ESR} &\tabincell{l}{SCAFFOLD \\ ($M=2$) } &1952$(1\times)$  &$\boldsymbol{}$0.4213  &$79.66\%$  &956$(1\times)$  & $\boldsymbol{}$0.3765 &$82.75\%$  &617$(1\times)$  &$\boldsymbol{}$0.3480  &$83.79\%$  \\ \cline{2-11} 
		&\tabincell{l}{SCAFFOLD \\ ($M=3$) }  &1972$(1\times)$  &$\boldsymbol{}$0.4208  &$79.75\%$  &944$(1\times)$  &$\boldsymbol{}$0.3767  &$82.49\%$  &619$(1\times)$  &$\boldsymbol{}$0.3470  &$84.05\%$   \\ \cline{2-11} 
		&\tabincell{l}{SCAFFOLD \\ ($M=5$) } &1941$(1\times)$  &$\boldsymbol{}$0.4212  &$79.79\%$  &969$(1\times)$  &$\boldsymbol{}$0.3765  &$82.62\%$  &621$(1\times)$  &$\boldsymbol{}$0.3485  &$84.07\%$   \\ \cline{2-11} 
		&\tabincell{l}{LoSAC \\ ($M=2$) }  &1020$\boldsymbol{(1.91\times)}$  &$\boldsymbol{0.3807}$  &$\boldsymbol{82.40}\%$  &498$\boldsymbol{(1.92\times)}$  &$\boldsymbol{0.3304}$  &$\boldsymbol{85.02}\%$  &328$\boldsymbol{(1.88\times)}$  &$\boldsymbol{0.2952}$  &$\boldsymbol{87.37}\%$ \\ \cline{2-11} 
		&\tabincell{l}{LoSAC \\ ($M=3$) }   &696$\boldsymbol{(2.83\times)}$  &$\boldsymbol{0.3561}$  & $\boldsymbol{83.13}\%$ &341$\boldsymbol{(2.77\times)}$  &$\boldsymbol{0.2931}$  &$\boldsymbol{87.30}\%$  &243$\boldsymbol{(2.55\times)}$  &$\boldsymbol{0.2503}$  &$\boldsymbol{89.28}\%$  \\ \cline{2-11} 
		&\tabincell{l}{LoSAC \\ ($M=5$) }  &445$\boldsymbol{(4.36\times)}$  &$\boldsymbol{0.3197}$  &$\boldsymbol{85.54}\%$  &206$\boldsymbol{(4.70\times)}$  &$\boldsymbol{0.2367}$  &$\boldsymbol{90.05}\%$  &180$\boldsymbol{(3.45\times)}$  &$\boldsymbol{0.1967}$  &$\boldsymbol{91.54}\%$  \\ \hline
	\end{tabular}
\end{table*}
\normalsize

\subsubsection{Compared algorithms} 
We compare LoSAC with five representative baseline algorithms 
FedAvg~\cite{pmlr-v54-mcmahan17a}, FedCM \cite{fedcm}, SCAFFOLD  \cite{SCAFFOLD}, FedADMM \cite{fedadmm}
and MimeSVRG~\cite{MIME}. {Moreover, to show the ineffectiveness of the naive extension of SAGA to FedOpt, we have also implemented FedSaga.} 
As is mentioned, FedAvg improves the communication efficiency comparing to FedSGD~\cite{pmlr-v54-mcmahan17a}. FedCM \cite{fedcm} adopts the momentum strategy in FedOpt. SCAFFOLD~\cite{SCAFFOLD} equips with the capability in handling non-IID data. {  Particularly, MimeSVRG is developed by adapting SVRG~\cite{NIPS2013_4937} to FedOpt using the framework Mime~\cite{MIME}. Especially, FedADMM  adapts the alternating direction method of multipliers (ADMM) to FedOpt \cite{fedadmm}, which are known to conveniently and efficiently solve the nonsmooth optimization problems \cite{admm}. 
	
			%
			%
			%
			%
			%
			%
			%
			%
			%
			%
			%
			%
			%
			%
			%
			%
			%

	For the default parameters for all algorithms, $(N,S)=(1,000,50)$ for MNIST cases and $(N,S)=(100,10)$ for HAR and ESR cases. 
	We set the local data division $M=5$ and the local iteration $T=5$.  
	Particularly, the step size is set to yield as the best performance as possible for each algorithm, 
	i.e., $\eta=4\times10^{-4}$ for MNIST cases and $\eta=10^{-4}$ for HAR and ESR cases. As for FedADMM \cite{fedadmm}, the details of FedADMM solving low rank matrix estimation are in Appendix \ref{app:fedadmm}. Different from \cite{pmlr-v54-mcmahan17a} that the local update in FedAvg traverses all the dataset for multiple epochs,  we follow \cite{SCAFFOLD} that all algorithms are implemented by sampling a mini-batch of data samples for search direction in the local update.

			%
				%
				%
				%
				%
				%
				%
				%
				%
				%
	
	\subsubsection{Evaluation tasks} 
	For the overall performance and ablation study, the cross entropy and the classification accuracy are evaluated. 	Specifically,	the overall performance is to show the general performance with different data and parameter settings, with the comparison to the baseline algorithms; the ablation study aims to show the communication and the computation efficiency of the algorithms for reaching a specific accuracy.  {Moreover, for the IID setting, the datasets are shuffled, and for the non-IID setting, the datasets are sorted by the labels. Then the datasets are divided evenly into $N$ clients.}   {Note since MimeSVRG's performance is comparative to SCAFFOLD and its computational complexity is twice of SCAFFOLD and LoSAC, hence  for the parameter sensitivity study, it is only implemented with MNIST.}  For DLG study, we mainly consider the similarity measured by the Frobenius norm between the estimated data samples by DLG and the real data samples. For the low rank matrix estimation, FedADMM \cite{fedadmm} and the proximal versions of LoSAC (LoSAC-Prox) and SCAFFOLD (SCAFFOLD-Prox) are implemented with the evaluation of the recovery matrix error and the recovery matrix rank.
	
	\begin{figure*} 
		\begin{minipage}[t]{0.15\linewidth} 
			\centering
			\begin{subfigure}{1.03\textwidth}
				\centering
				\includegraphics[width=1.0\linewidth]{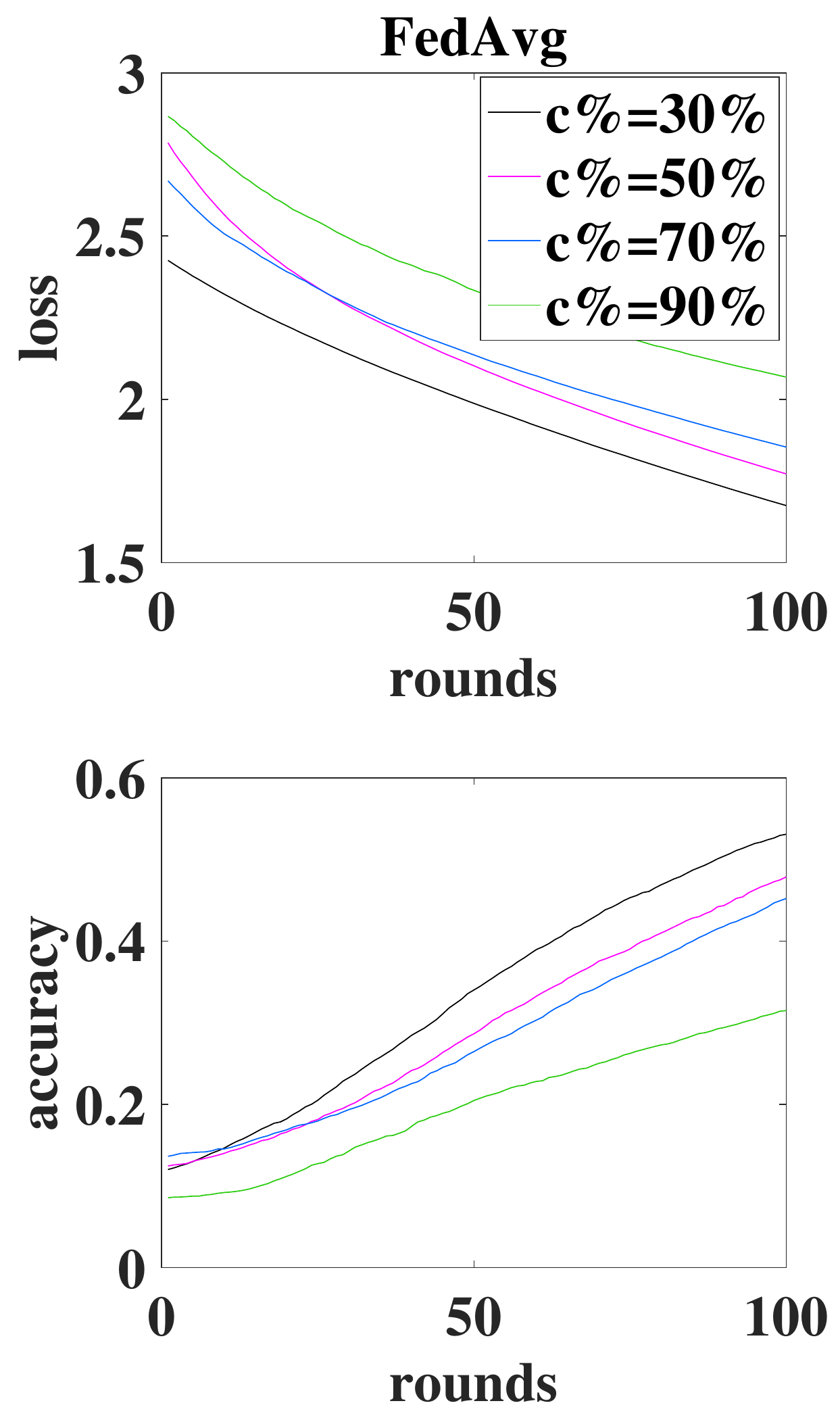}
				\label{fig:har_s10}
			\end{subfigure}\\
		\end{minipage} \hfill 
		\begin{minipage}[t]{0.15\linewidth} 
			\centering
			\begin{subfigure}{1.03\textwidth}
				\centering
				\includegraphics[width=1.0\linewidth]{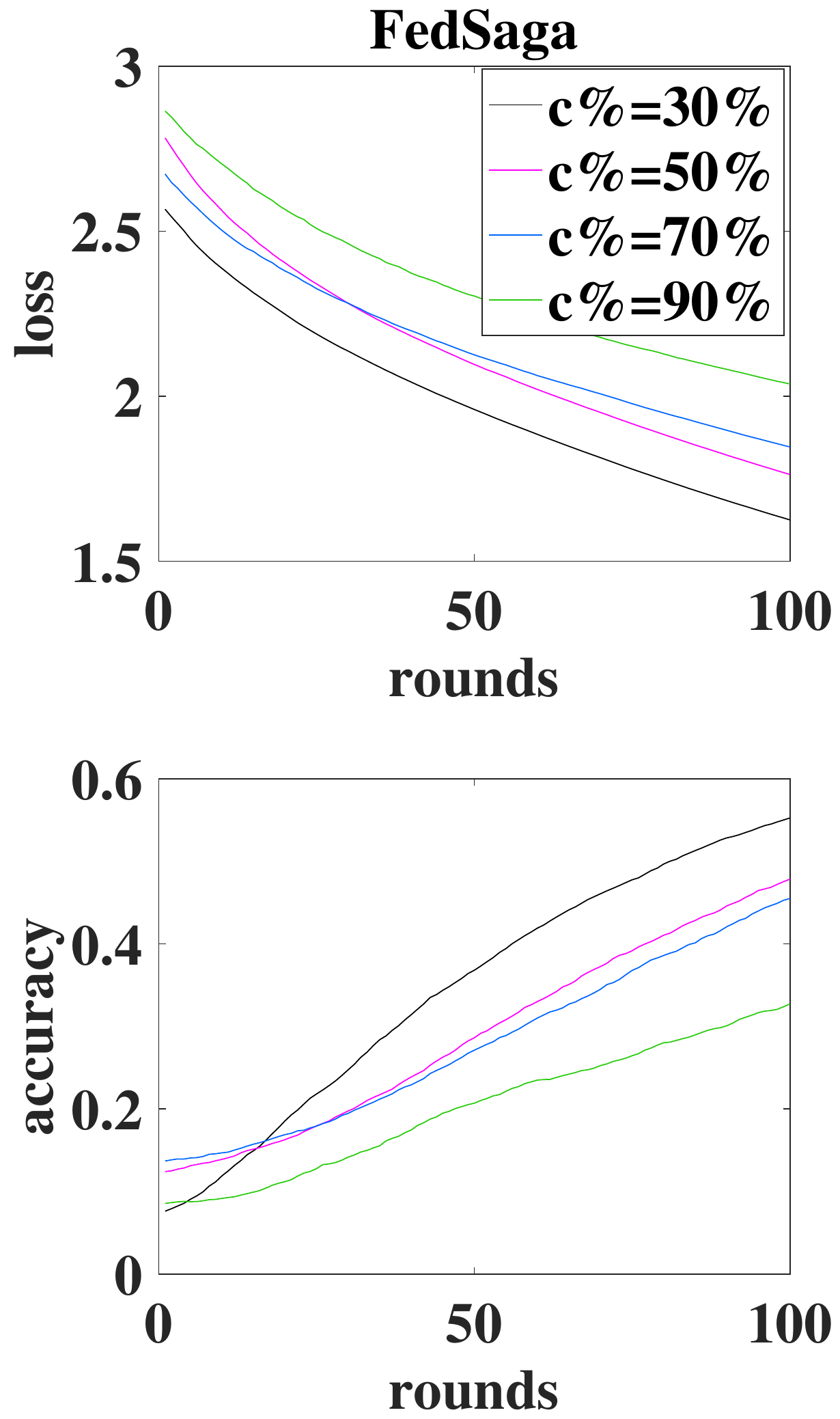}
				\label{fig:har_s30a}
			\end{subfigure}\\
			\label{fig:har_s30b}
		\end{minipage} \hfill
	\begin{minipage}[t]{0.15\linewidth} 
		\centering
		\begin{subfigure}{1.03\textwidth}
			\centering
			\includegraphics[width=1.0\linewidth]{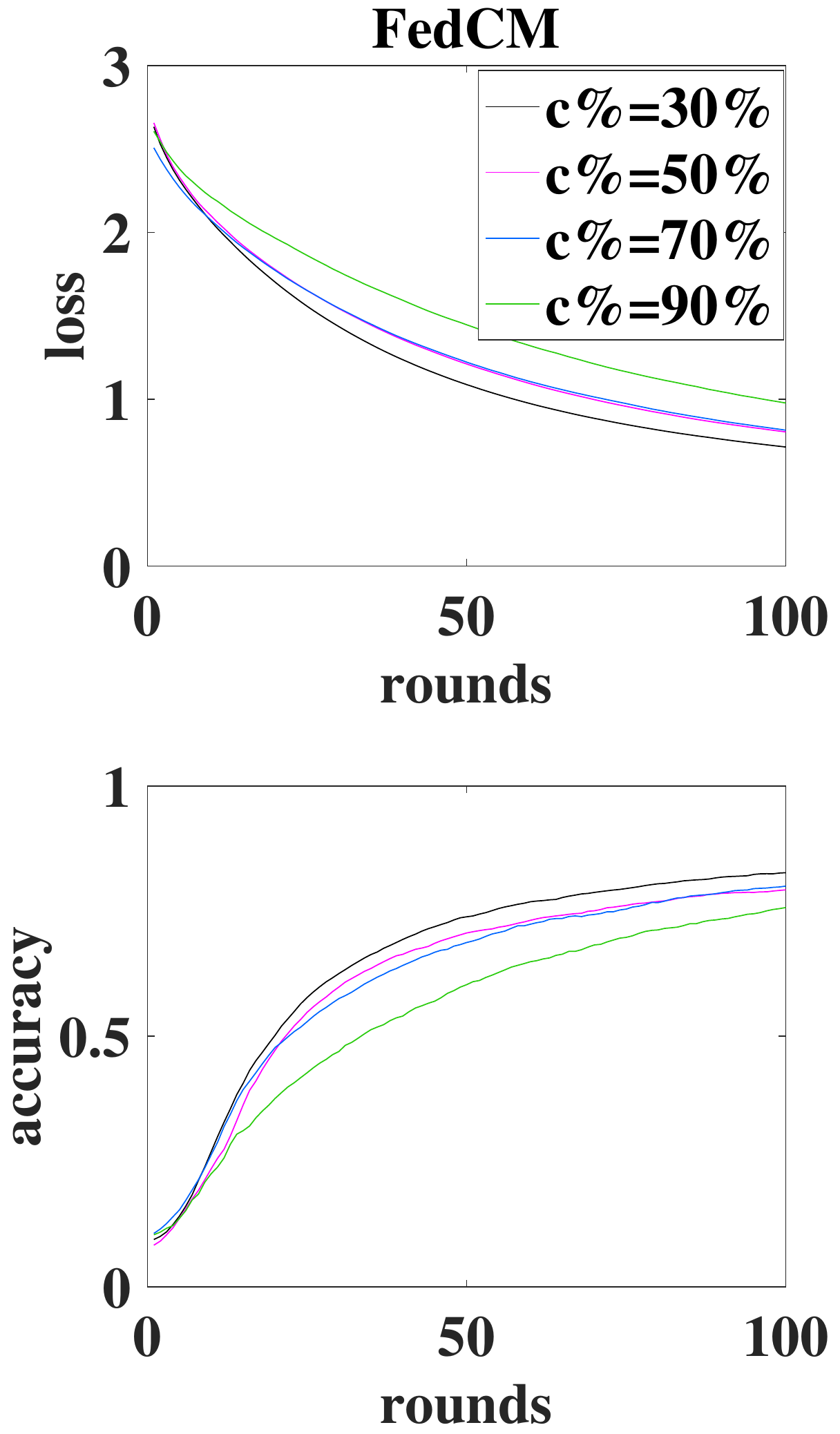}
			\label{fig:har_s30a}
		\end{subfigure}\\
		\label{fig:har_s30b}
	\end{minipage} \hfill
		\begin{minipage}[t]{0.15\linewidth} 
			\centering
			\begin{subfigure}{1.03\textwidth}
				\centering
				\includegraphics[width=1.0\linewidth]{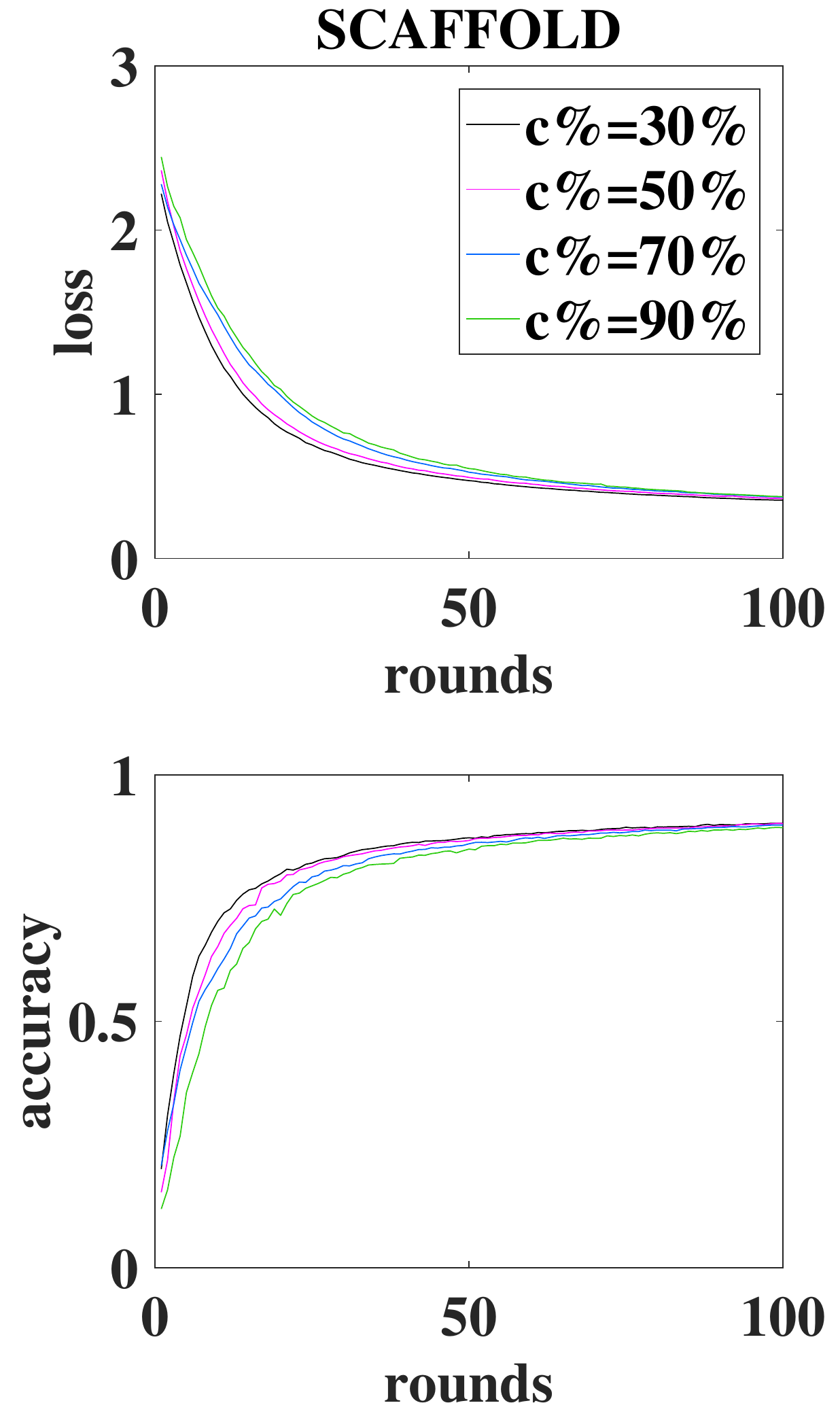}
				\label{fig:har_s50a}
			\end{subfigure}\\
			\label{fig:har_s50b}
		\end{minipage} \hfill
		\begin{minipage}[t]{0.15\linewidth} 
			\centering
			\begin{subfigure}{1.03\textwidth}
				\centering
				\includegraphics[width=1.0\linewidth]{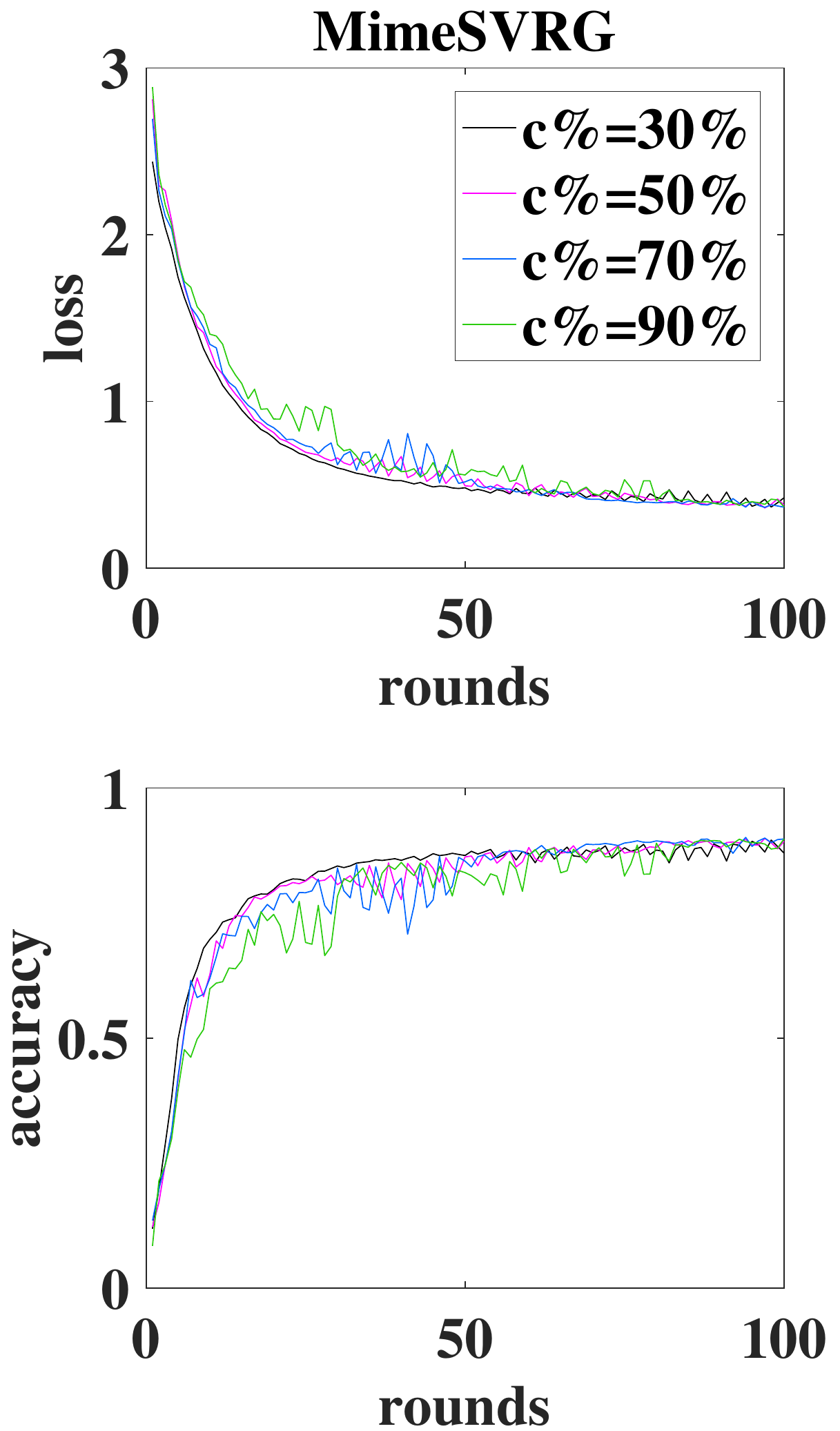}
				\label{fig:har_s100a}
			\end{subfigure}\\
			\label{fig:har_s50b}
		\end{minipage} \hfill
		\begin{minipage}[t]{0.15\linewidth} 
			\centering
			\begin{subfigure}{1.03\textwidth}
				\centering
				\includegraphics[width=1.0\linewidth]{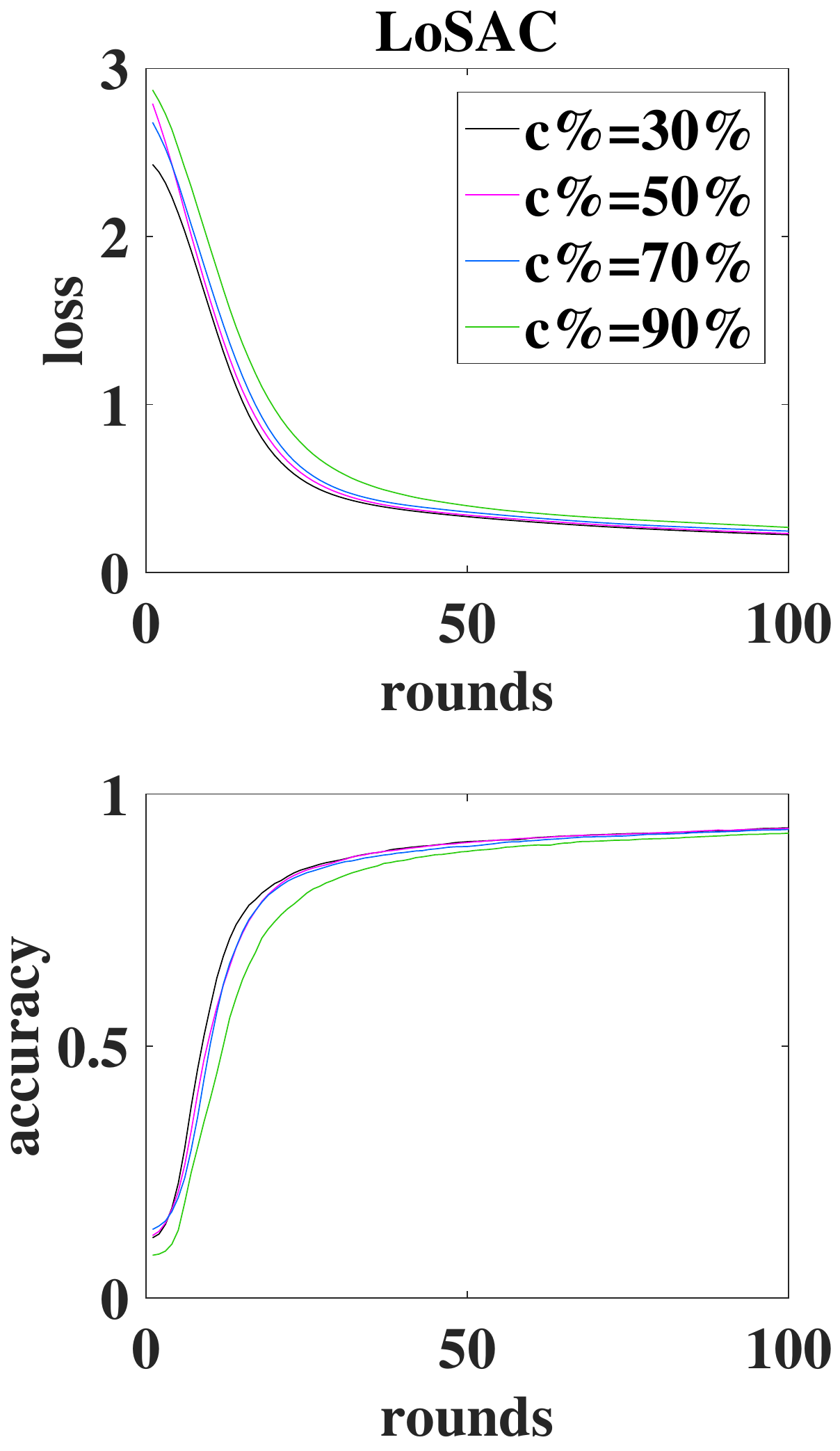}
				\label{fig:har_s50a}
			\end{subfigure}\\
			\label{fig:har_s50b}
		\end{minipage} 
		
		\caption{The performance evaluations of all algorithms with different data heterogeneity parameter $c\%$ (larger $c\%$ means the dataset is more heterogenous).}
		\label{fig:har}
	\end{figure*}
	
	\subsection{Overall Performance}
	\subsubsection{Data Heterogeneity}
	{In this subsection, we study the effect of the data heterogeneity on LoSAC and show the impacts of the local and global variance reduction schemes. Specifically, FedSaga utilizes local variance reduction, while SCAFFOLD, MimeSVRG and LoSAC uses global one. Moreover, FedAvg is implemented as the benchmark.  For the experimental settings, the MNIST dataset is utilized for training and testing. We choose $(1-c\%)$ of the uniformly shuffled training dataset and distribute it to $N=1,000$ clients. Then, the left $c\%$ of the training dataset is set to be non-IID, namely is sorted by the labels and distributed to all clients. Therefore, the larger the parameter $c\%$ is, the data is more heterogeneous. For other parameters settings, $S$ is set to $S=50$, the step size is chosen as $\eta=4\times 10^{-4}$ for all algorithms since it has shown the best performances. Moreover, the iteration number is $(R,T)=(100,5)$. For LoSAC and FedSaga, Since the local dataset is divided into $M$ blocks, the parameter is set as $M=5$. }
	
	{As Fig. 2 shows,  when the data is more heterogenous (namely $c\%$ is larger), both FedAvg and FedSaga suffers from the data heterogeneity problem more seriously. For SCAFFOLD, MimeSVRG and LoSAC, since they all utilize the global information to correct the bias from the global model in the local update, the data heterogeneity problem has the little impact on the performance. Hence, the global variance reduction is much more robust to the data heterogeneity problem than the local one. Moreover, it can be seen that FedCM also has the impact of data heterogeneity, but the impact
		is not as serious as FedAvg and FedSaga. This can be attributed to the global aggregation of the
		local momentum term in FedCM, which has mitigated the model divergence problem.}

	\subsubsection{Parameter Sensitivity}
	We conduct extensive experiments to study the effects of $\{T,S\}$, which play significant roles on the convergence results in Theorem \ref{theorem1}. The experiments use IID and non-IID settings for each evaluation respectively. The results are shown in Figs. 3$\sim$4, Fig. 5 and Fig. 6 with MNIST, HAR and ESR datasets respectively. In particular,  {HAR} and {ESR} datasets correspond to the mobile and the medical applications respectively. While we uses the default settings for MNIST, we fix $N=100$, $\eta=10^{-4}$ and $M=5$ for HAR and ESR. In general, larger $T$ and $S$ lead to faster speed and better classification performance for LoSAC. This matches well with the convergence results in Theorem \ref{theorem1}.

	\begin{figure*} 
		\begin{minipage}[t]{0.24\linewidth} 
			\centering
			\begin{subfigure}{1.02\textwidth}
				\centering
				\includegraphics[width=1.0\linewidth]{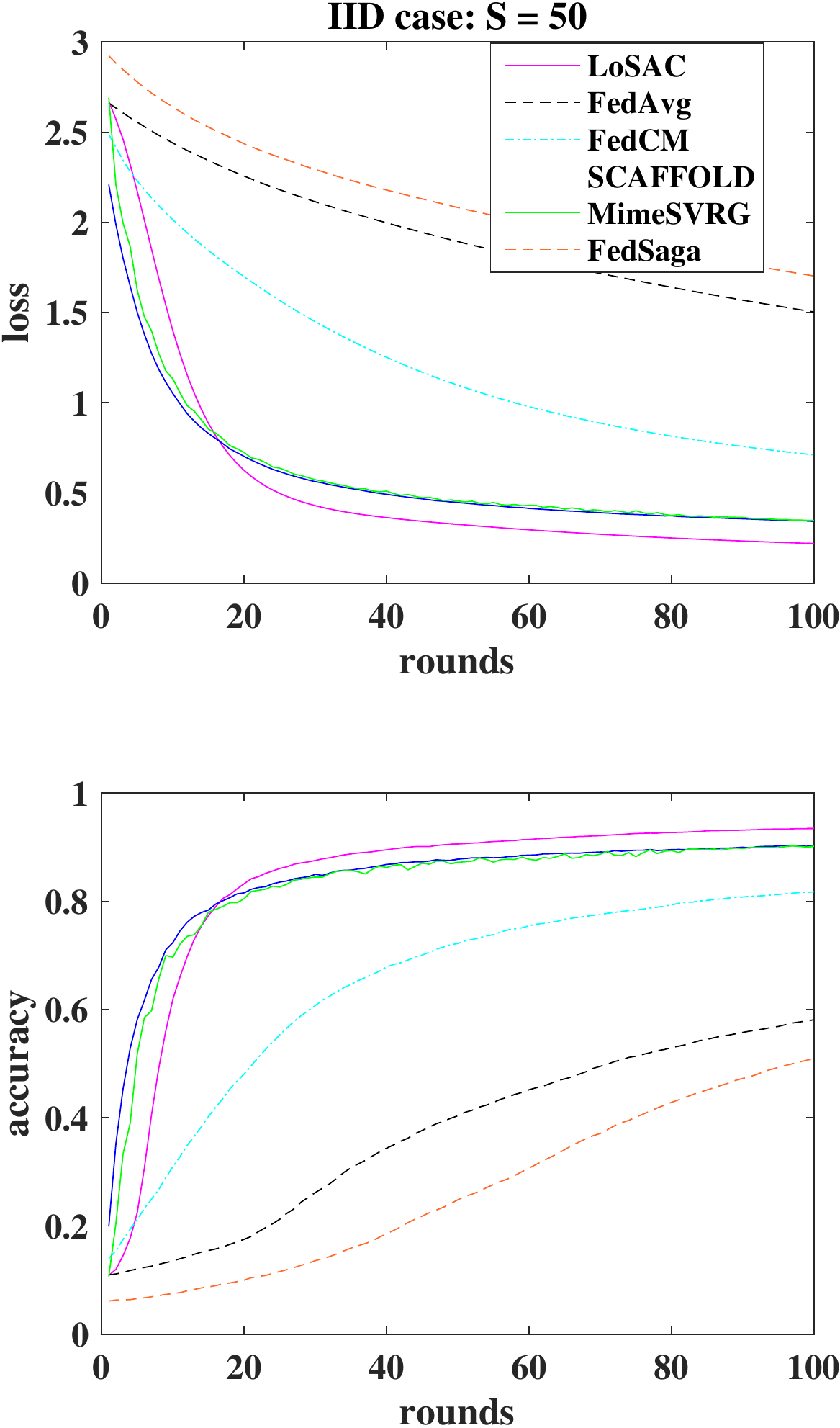}
				\label{fig:har_s10}
			\end{subfigure}\\
		\end{minipage} \hfill 
		\begin{minipage}[t]{0.24\linewidth} 
			\centering
			\begin{subfigure}{1.03\textwidth}
				\centering
				\includegraphics[width=1.0\linewidth]{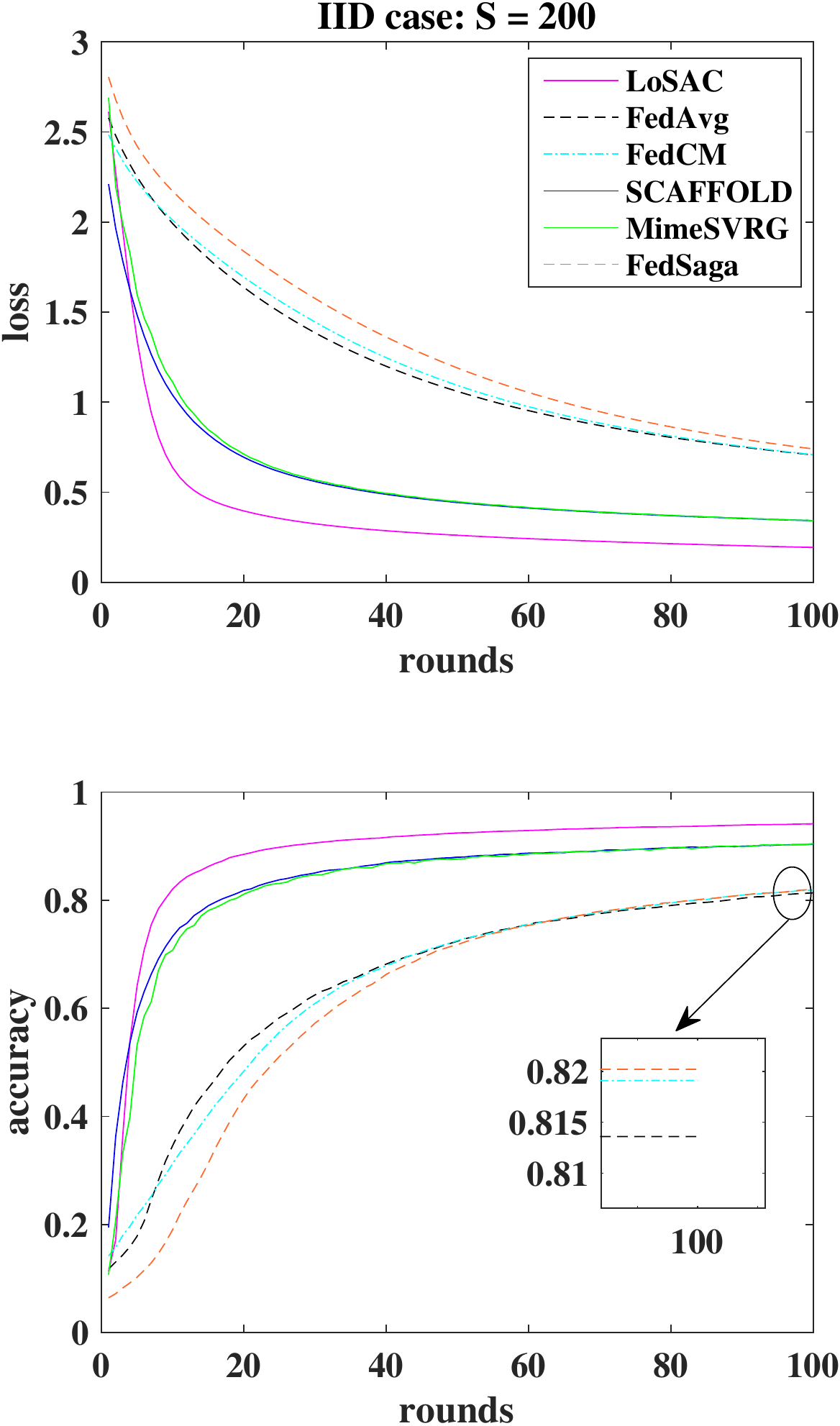}
				\label{fig:har_s30a}
			\end{subfigure}\\
			\label{fig:har_s30b}
		\end{minipage} \hfill
		\begin{minipage}[t]{0.24\linewidth} 
			\centering
			\begin{subfigure}{1.02\textwidth}
				\centering
				\includegraphics[width=1.0\linewidth]{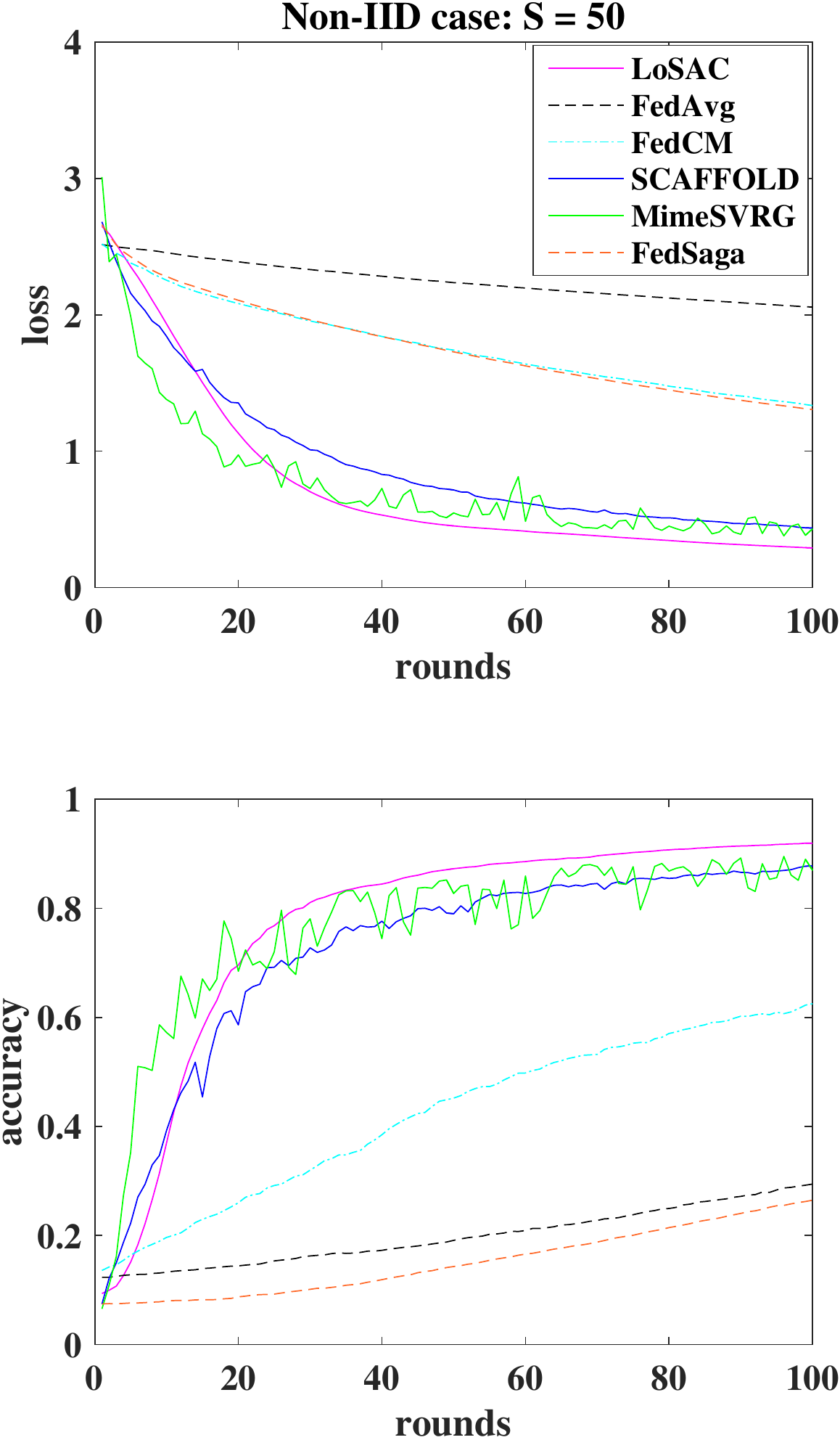}
				\label{fig:har_s50a}
			\end{subfigure}\\
			\label{fig:har_s50b}
		\end{minipage} \hfill
		\begin{minipage}[t]{0.24\linewidth} 
			\centering
			\begin{subfigure}{1.04\textwidth}
				\centering
				\includegraphics[width=1.0\linewidth]{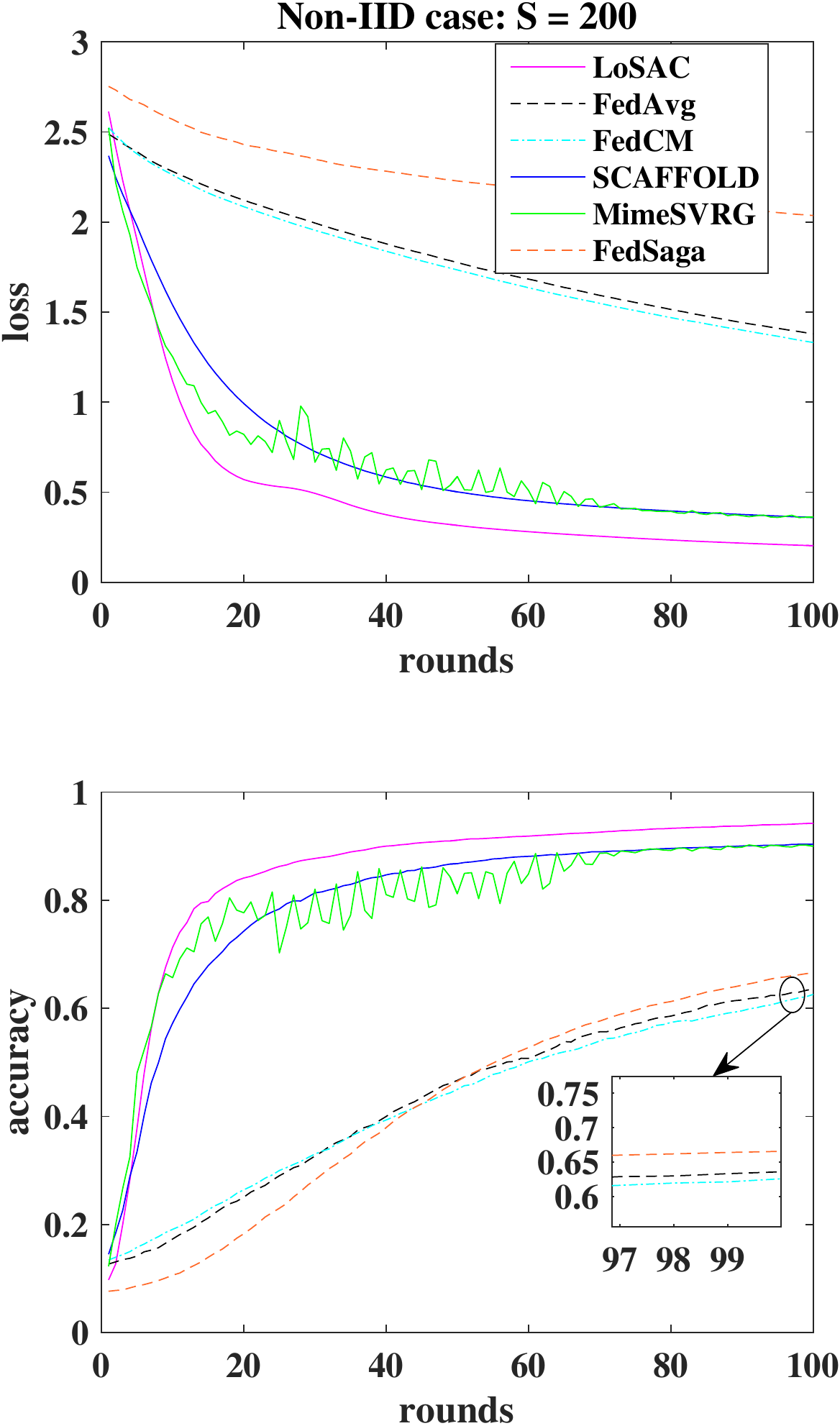}
				\label{fig:har_s100a}
			\end{subfigure}\\
			\label{fig:har_s50b}
		\end{minipage}%
		
		\caption{The performance evaluations with different client numbers using real dataset \textit{MNIST} in IID and non-IID settings.}
		\label{fig:har}
	\end{figure*}
	
	\begin{figure*} 
		\begin{minipage}[t]{0.24\linewidth} 
			\centering
			\begin{subfigure}{1.03\textwidth}
				\centering
				\includegraphics[width=1.0\linewidth]{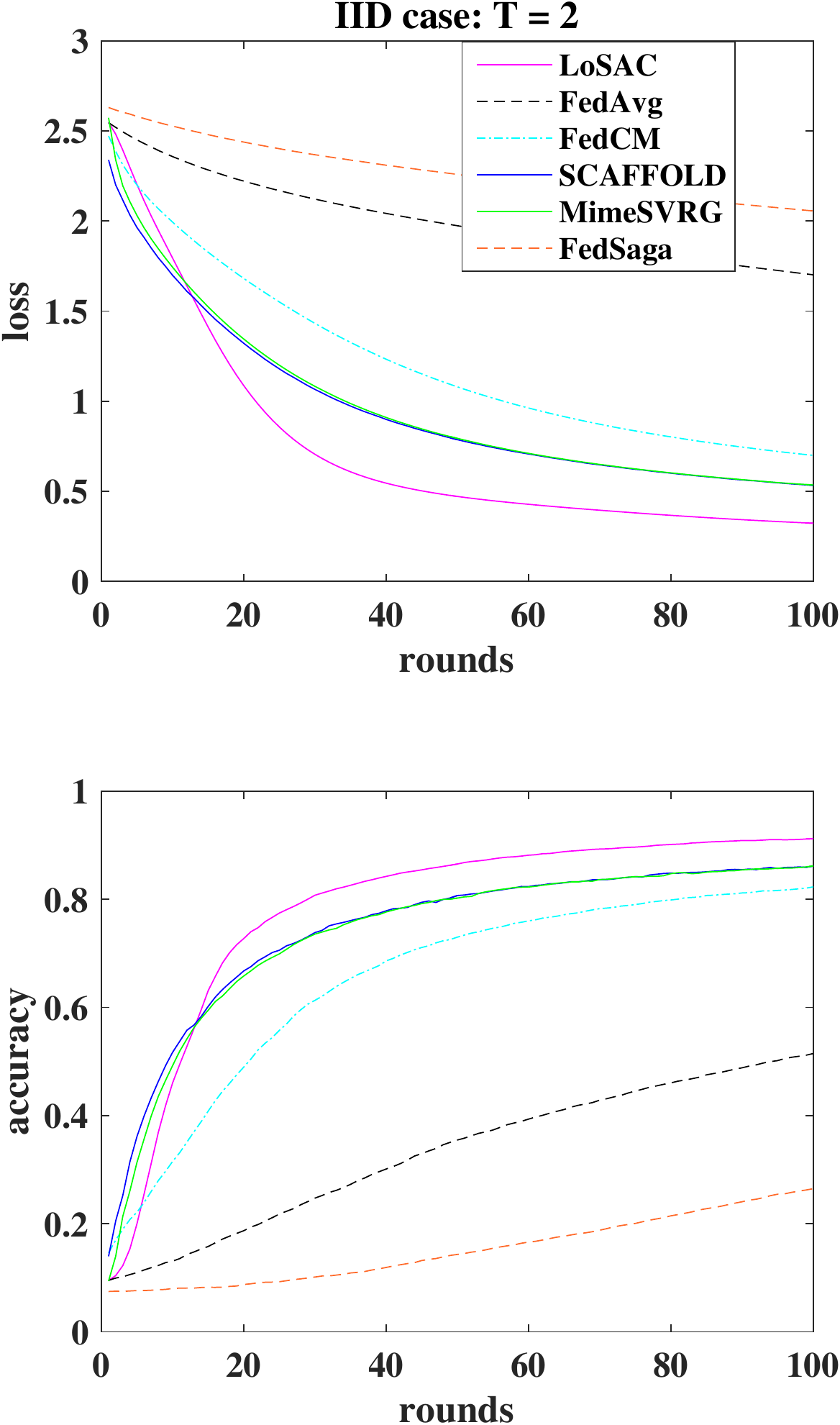}
				\label{fig:har_s10}
			\end{subfigure}\\
		\end{minipage} \hfill 
		\begin{minipage}[t]{0.24\linewidth} 
			\centering
			\begin{subfigure}{1.03\textwidth}
				\centering
				\includegraphics[width=1.0\linewidth]{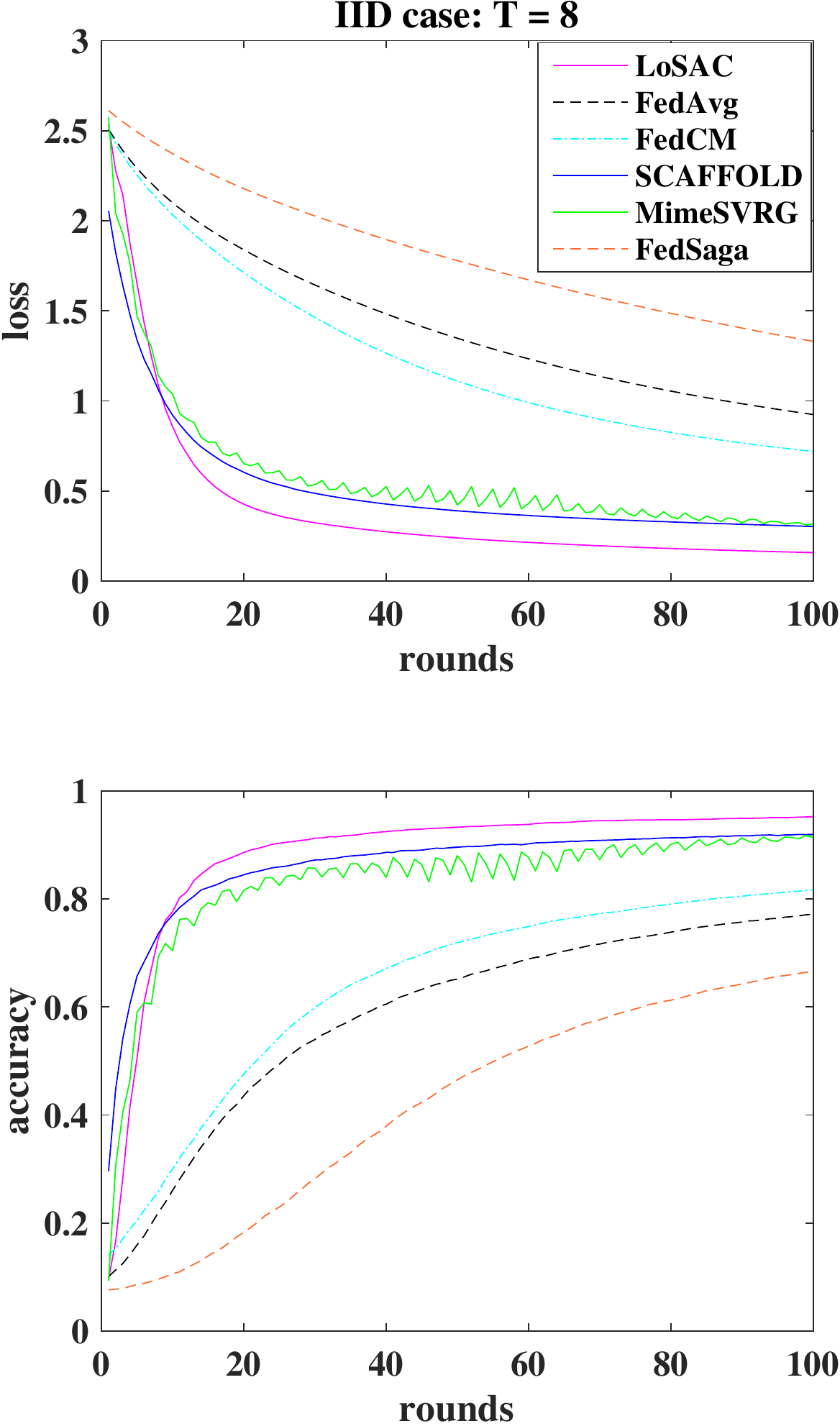}
				\label{fig:har_s30a}
			\end{subfigure}\\
			\label{fig:har_s30b}
		\end{minipage} \hfill
		\begin{minipage}[t]{0.24\linewidth} 
			\centering
			\begin{subfigure}{1.02\textwidth}
				\centering
				\includegraphics[width=1.0\linewidth]{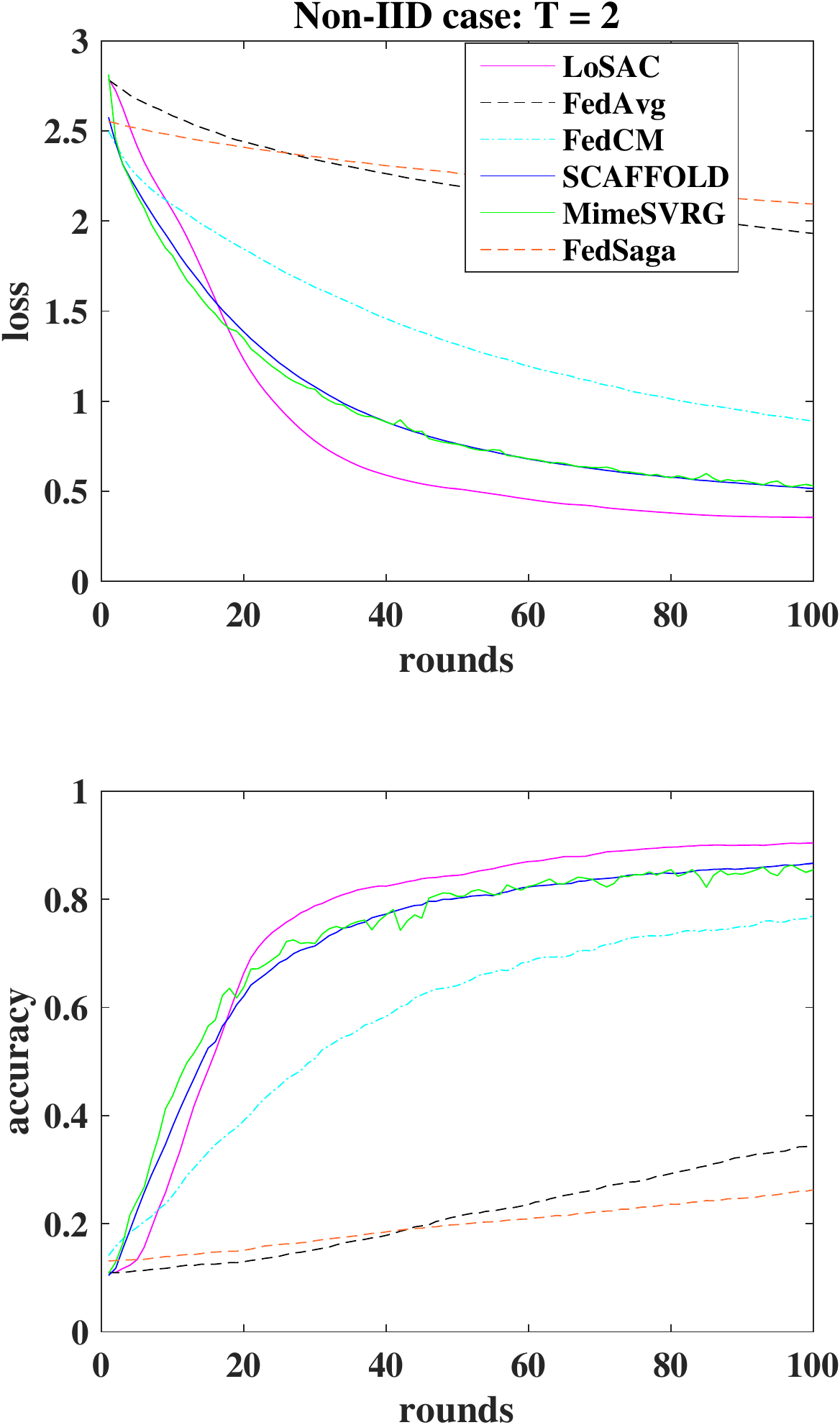}
				\label{fig:har_s50a}
			\end{subfigure}\\
			\label{fig:har_s50b}
		\end{minipage} \hfill
		\begin{minipage}[t]{0.24\linewidth} 
			\centering
			\begin{subfigure}{1.03\textwidth}
				\centering
				\includegraphics[width=1.0\linewidth]{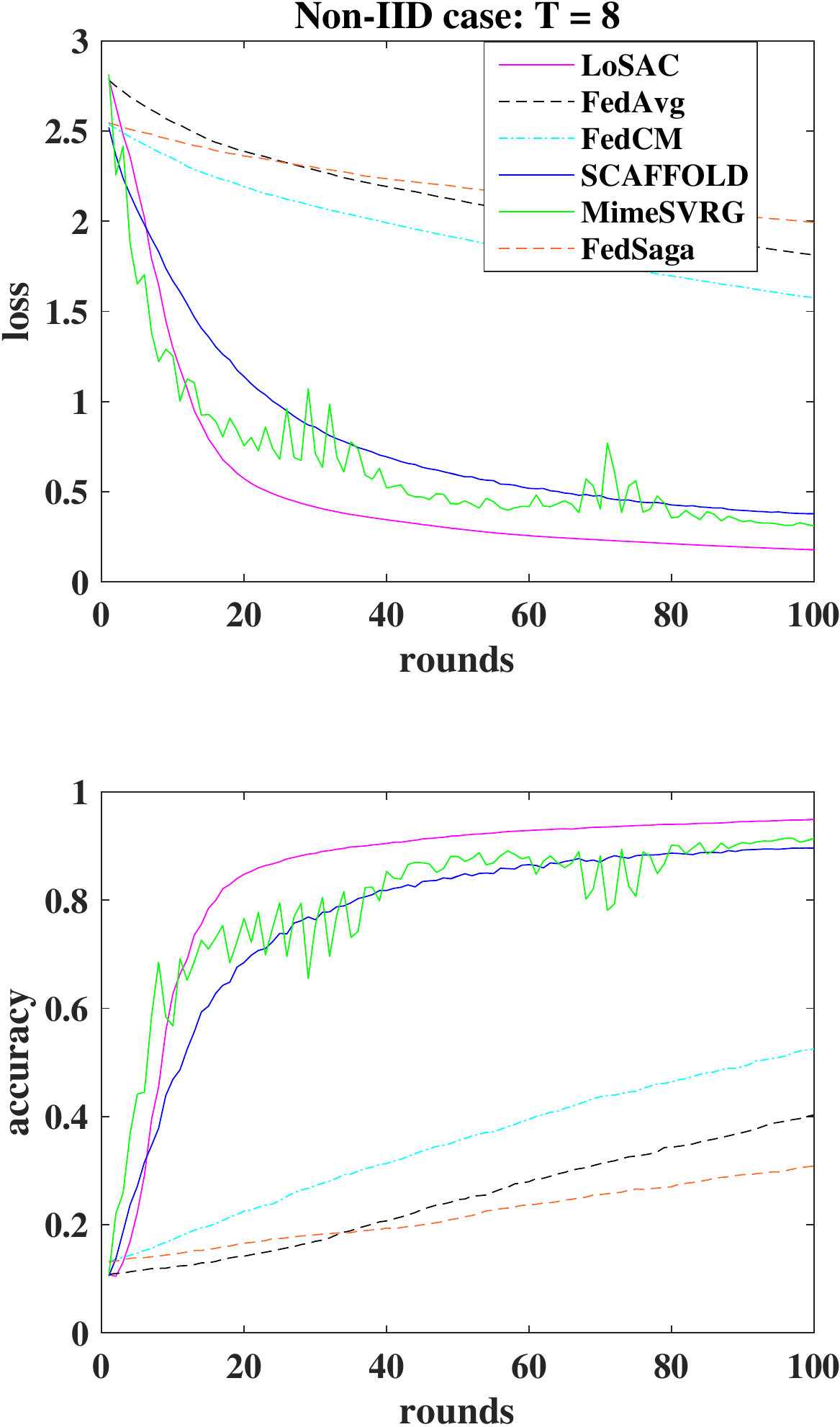}
				\label{fig:har_s100a}
			\end{subfigure}\\
			\label{fig:har_s50b}
		\end{minipage}%
		
		\caption{The performance evaluations with different local steps using real dataset \textit{MNIST} in IID and non-IID settings.}
		\label{fig:har}
	\end{figure*}

	{In Figs. 3$\sim$4, while our proposed method exhibits the prominent performance improvements, the
		non-IID data has significantly degraded the performances of FedAvg and MimeSVRG. With regard
		to the comparison of FedSaga and FedAvg, it can be seen that even with the acceleration scheme
		(here it is variance reduction), FedSaga performs worse than FedAvg. Although the acceleration
		scheme has shown the effectiveness in centralized optimization methods, it is not effective in the
		naive extension of SAGA to the FedOpt settings. This may due to the reason that the acceleration
		scheme using only local information may adversely lead FedSaga to fast approach the local optimum
		instead of the global one, resulting in large bias from the global optimum. For FedCM, more clients
		participated for local updates in FedCM can generally lead to the higher performances, this can be attributed to that more clients will contribute more information. Moreover, the non-IID case has
		affected the performance of FedCM, and more local steps in FedCM will adversely lead performance
		degradation. This is due to the reason that FedCM uses momentum acceleration in local model
		update, but with non-IID setting, it will accelerate the speed of model divergence and lead to
		performance degradation. In particular for MimeSVRG, it shows the large fluctuations in the
		non-IID setting, and more local iterations will lead to larger fluctuations. While SCAFFOLD and
		our method exhibit the strong capability in handling data heterogeneity problem, our method
		outperforms SCAFFOLD. Thus, it demonstrates the strong capability of LoSAC for mitigating the
		model divergence problem.}   
	
	For HAR, Fig. 5 has shown the performances of all methods with different $S$ and $T$. In general, the result matches the Theorem \ref{theorem1} that larger $T$ and $S$ leads to better performances. In particular,  our proposed method with $T=2$ even exhibits better performances than SCAFFOLD with $T=8$. This means that with only $25\%$  of the computational complexity in SCAFFOLD, our proposed method still yields quite high performances. {However, it shows the large fluctuations of LoSAC in the initial few updates. The reason may due to the randomness in the delayed full gradient that has brought the large variance in the initial updates,  when $z_{i,j}$ seriously differs from $x_i$. When the algorithm progresses, it is expected to satisfy $z_{i,j} \rightarrow x_i$, and the variance begins to reduce. }

	Fig. 5 has also exhibited the cases with ESR dataset using the same settings. Here, both SCAFFOLD and FedAvg have been significantly affected by the model divergence problem, while our proposed method has demonstrated the remarkable performances.  
	
	\begin{figure*} 
		\begin{minipage}[t]{0.24\linewidth} 
			\centering
			\begin{subfigure}{1.03\textwidth}
				\centering
				\includegraphics[width=1.0\linewidth]{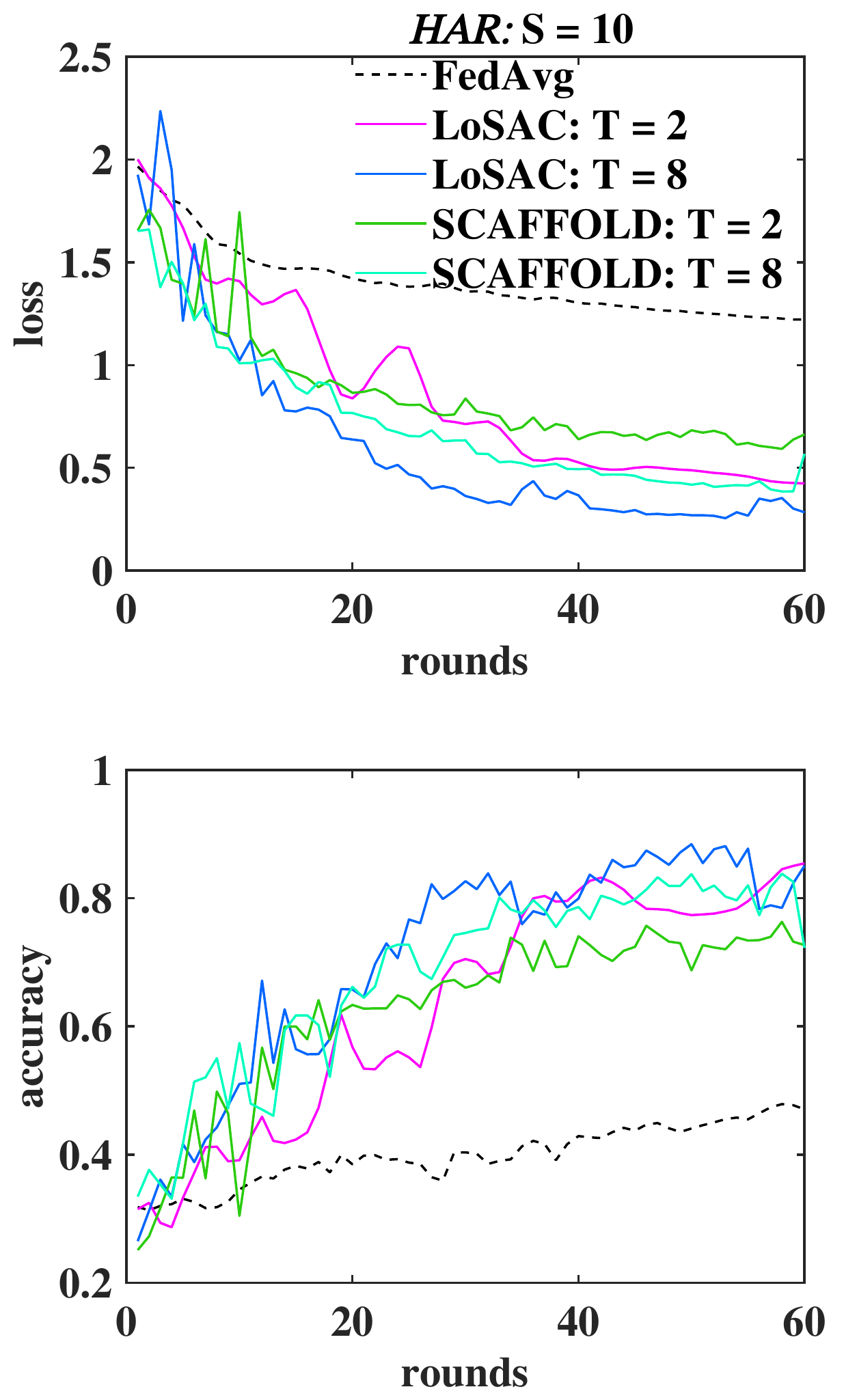}
				\label{fig:har_s10}
			\end{subfigure}\\
		\end{minipage} \hfill 
		\begin{minipage}[t]{0.24\linewidth} 
			\centering
			\begin{subfigure}{1.03\textwidth}
				\centering
				\includegraphics[width=1.0\linewidth]{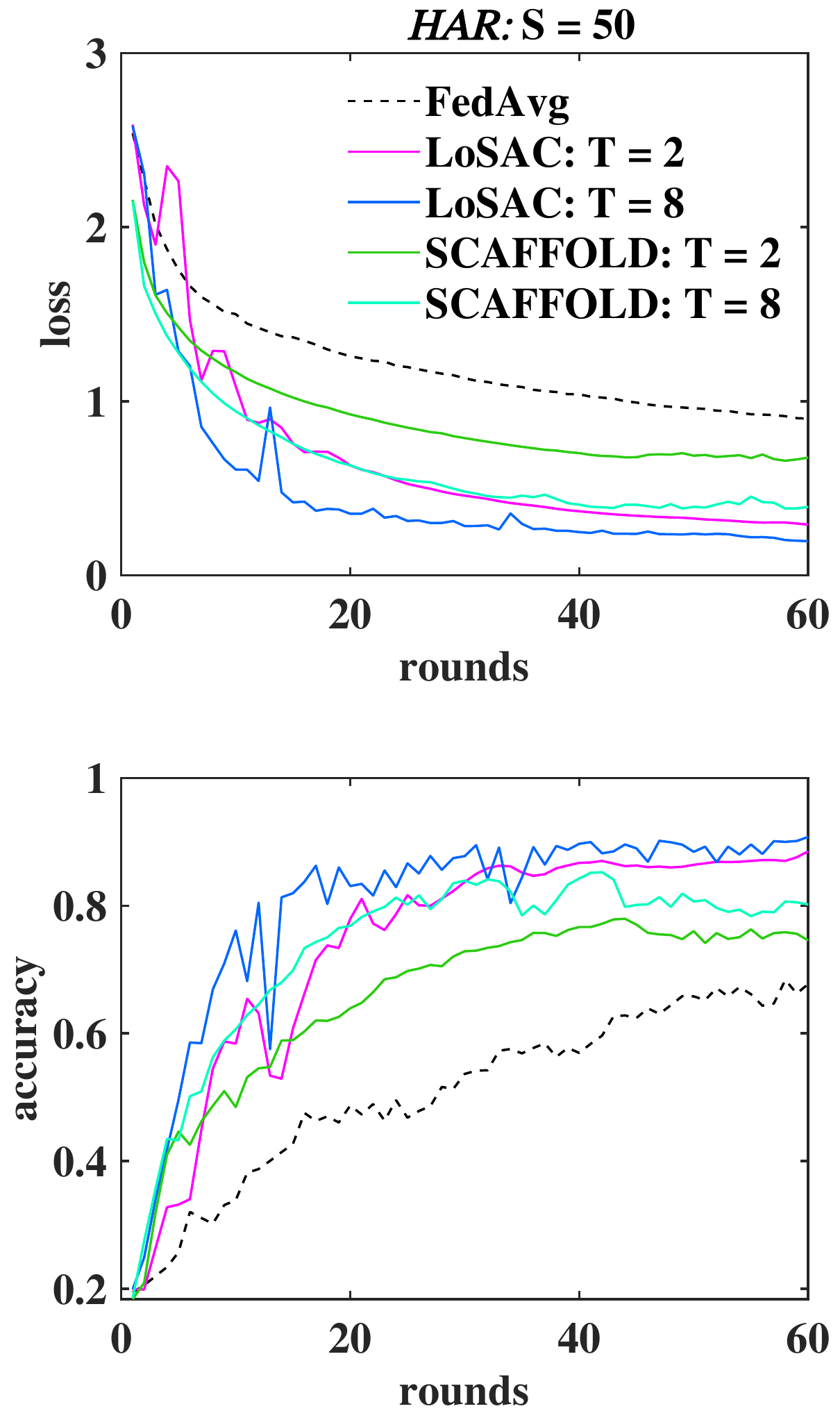}
				\label{fig:har_s30a}
			\end{subfigure}\\
			\label{fig:har_s30b}
		\end{minipage} \hfill
		\begin{minipage}[t]{0.24\linewidth} 
			\centering
			\begin{subfigure}{1.02\textwidth}
				\centering
				\includegraphics[width=1.0\linewidth]{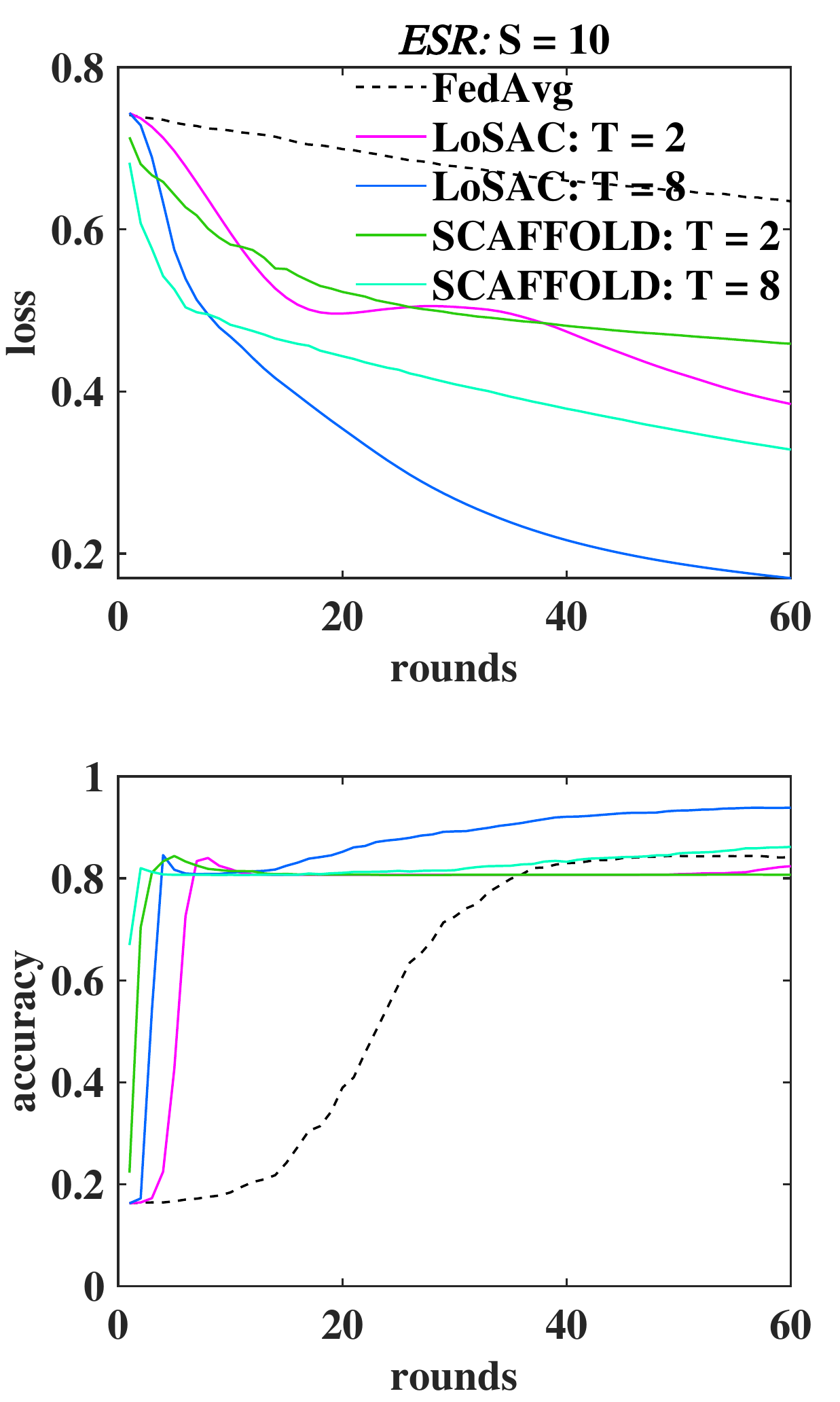}
				\label{fig:har_s50a}
			\end{subfigure}\\
			\label{fig:har_s50b}
		\end{minipage} \hfill
		\begin{minipage}[t]{0.24\linewidth} 
			\centering
			\begin{subfigure}{1.03\textwidth}
				\centering
				\includegraphics[width=1.0\linewidth]{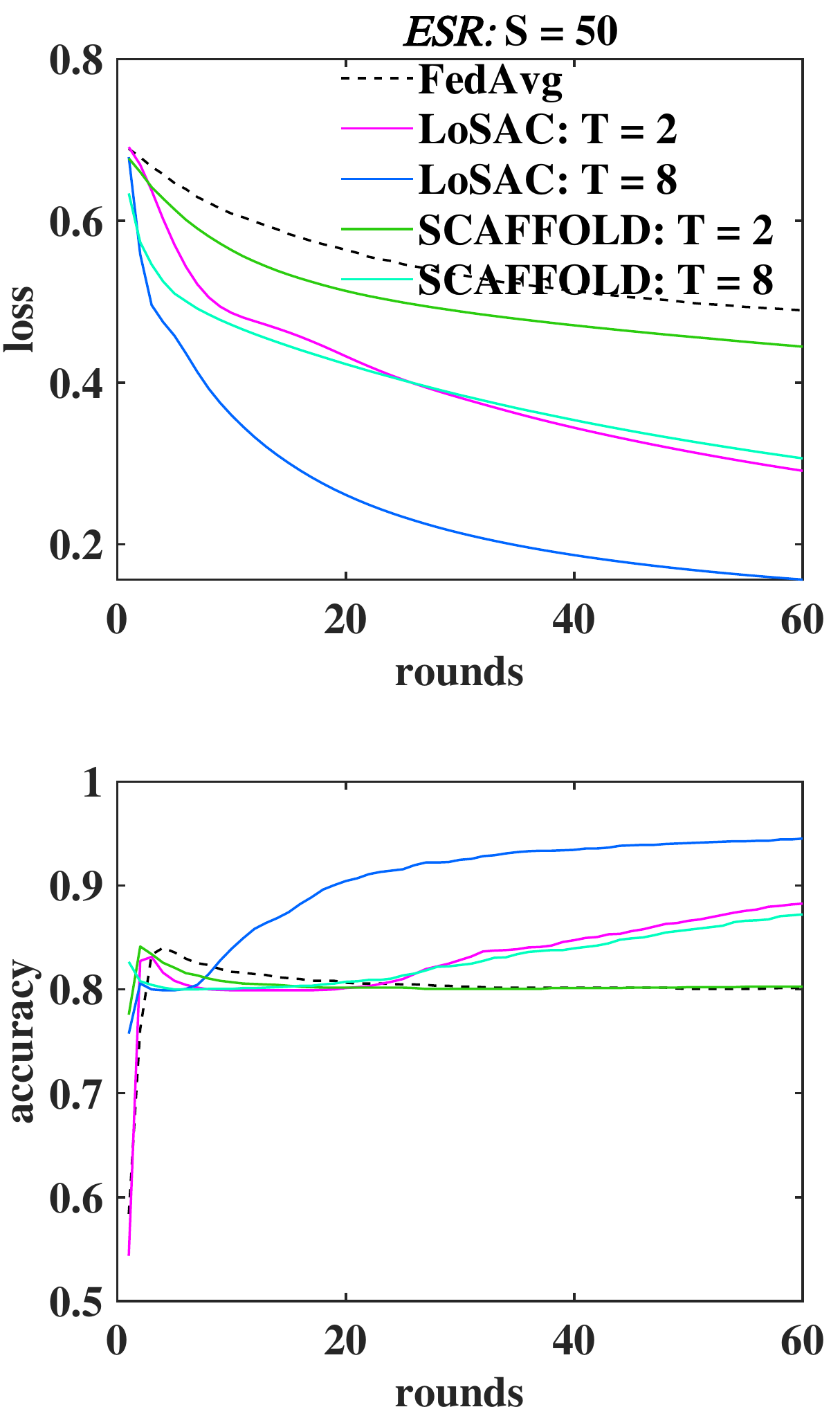}
				\label{fig:har_s100a}
			\end{subfigure}\\
			\label{fig:har_s50b}
		\end{minipage}%
		
		\caption{The performance evaluations with different client number $S$ and local iteration number $T$ using real datasets \textit{HAR} and \textit{ESR} in non-IID setting, figures in the above row are loss function evaluations and the below figures are evaluations of classification accuracy.}
		\label{fig:har}
	\end{figure*}

	\subsection{Ablation Study}
	We have shown the overall performance with different parameter and data heterogeneity settings. In this subsection, we continue to conduct the ablation study with different local memory (corresponding to the local data division $M$) and local iterations. We implement SCAFFOLD as the benchmark since it performs the second best in overall performance. Moroever, we adopt the non-IID setting. The step size for all cases is set to $\eta=10^{-4}$ for the algorithms to yield as the best performances as possible. We set $(N,S)=(100,10)$ for all cases. $\{T,M\}$ are tuned to for the comparisons. In general, it shows the significantly higher communication and computation efficiency over SCAFFOLD, which also demonstrates the effectiveness of the estimate for the global full gradient. The results are shown in Table \ref{table_datasets}.

	\subsubsection{Communication efficiency}
	It can be seen from 
	Table~\ref{table_datasets} that LoSAC requires much fewer communication rounds  than  SCAFFOLD to reach a given accuracy. 
	In particular, the communication efficiency is improved by  more than $300\%$ for ESR case and $100\%$ in average { for all cases.}  
	Thus, LoSAC is  communication quite efficient. 
	{This is because LoSAC estimates the global gradient more accurately than SCAFFOLD,} 
	which has accelerated the convergence speed and mitigated the model divergence problem.
	
	\subsubsection{Computation efficiency}
	Table~\ref{table_datasets} further demonstrates the high computation efficiency of LoSAC. To be specific, when LoSAC with $T=2$ and $M=3$, it  requires comparable communication rounds to reach the specific accuracy with SCAFFOLD with $T=6$.  This means with only around $33\%$ computation complexity of SCAFFOLD, LoSAC can still yield higher communication efficiency. 
	
	\subsubsection{Local memory.}  For LoSAC, each client $i$ needs to spend the sufficient memory to store $y_{i,j}, j\in[M]$, depending on the partitioning of the local dataset $\mathcal{D}_i$. Hence, we choose different memory sizes for the evaluations, i.e., $M=\{2,3,5\}$. It also corresponds to $M$ divisions of the local datasets. {Table~\ref{table_datasets} indicates that larger memory size leads to a better performance for LoSAC.} This may due to the reason that the local estimation of the global full gradient is improved with a larger memory size, and thus is less affected by the non-IID data. However, it will also cost local resources. In FL applications, clients may have limited memory, e.g., the case for mobile phones, thus we suggest a better trade-off between the performance and the resource. Note when $M=2$, the storage cost of LoSAC is $\mathcal{O}(2d)$ and the same with  SCAFFOLD (SCAFFOLD requires to store the control variates and the model parameter for each local iteration), LoSAC  yields much better performances than  SCAFFOLD, i.e., more than $40\%$ averaged performance improvements. 
	
	\subsection{Defense Against DLG}
	We study the defense ability of each algorithm against DLG in Figure \ref{fig:dlg}. Specifically, except for LoSAC, SCAFFOLD, FedAvg and MimeSVRG,  we also implement DSGD since it has been the major attack objective by DLG \cite{44634}).  The real datasets MNIST, HAR and ESR are utilized for the evaluations. For simplicity, we perform each FedOpt on logistic regression for binary classification on all datasets. Specifically, for MNIST, the label is set to $0$ when it is smaller or equal to $5$ and $1$ otherwise; for HAR, the label is  set to $0$ when it is smaller or equal to $3$ and $1$ otherwise; For ESR, it has two classes that match well with binary logistic regression.  We set $(N,S)=(100,100)$, i.e., all the clients participate in the local update in each round. All local datasets are utilized for the search direction in each local step, i.e., $M=1$. Furthermore, we apply gradient descent (GD) to perform the DLG attack in Definition 1.   We tune the step size in the (a) MNIST  and HAR cases:  $\eta=10^{-4}$ for all algorithms and $\eta_d=10^{-3}$ for GD in DLG, (b) ESR case:   $\eta=10^{-3}$ for all algorithms and $\eta_d=10^{-3}$ for GD in DLG. For each case, the GD in DLG has the iteration number $100$. For simplicity without loss of generalization, we perform the DLG attack aiming to obtain the $1$st client's data samples $\mathcal{D}_1$ in the $5$th round for each algorithm in all cases. We denote $\hat{\mathcal{D}}_1$ as the obtained data samples by DLG, and use the metric $ \|\hat{\mathcal{D}}_1-\mathcal{D}_1\|_F$ for performance evaluations. Therefore, smaller value of the metric  $ \|\hat{\mathcal{D}}_1-\mathcal{D}_1\|_F$ means more successful for DLG to obtain the local dataset  $\mathcal{D}_1$.
	
	As shown in Fig. \ref{fig:dlg}, both DSGD and MimeSVRG are vulnerable to the DLG since the estimated data samples are more and more approaching to the true data samples. This is because they transmits the gradients and the corresponding models, which satisfies exactly the DLG attack in Definition 1. As for FedAvg, SCAFFOLD and LoSAC, the DLG aims to obtain the data samples based on the term $\nicefrac{1}{\eta L}\cdot (x^+_1-x)$, which is essentially the averaged gradients over the $L$ local steps. Under DLG attacks, the estimated data samples in FedAvg, SCAFFOLD and LoSAC are more and more divergent from the true data samples, which demonstrates the capabilities of these algorithms to defense against DLG.

	\begin{figure*} 
		\begin{minipage}[t]{0.32\linewidth} 
			\centering
			\begin{subfigure}{1\textwidth}
				\centering
				\includegraphics[width=1.0\linewidth]{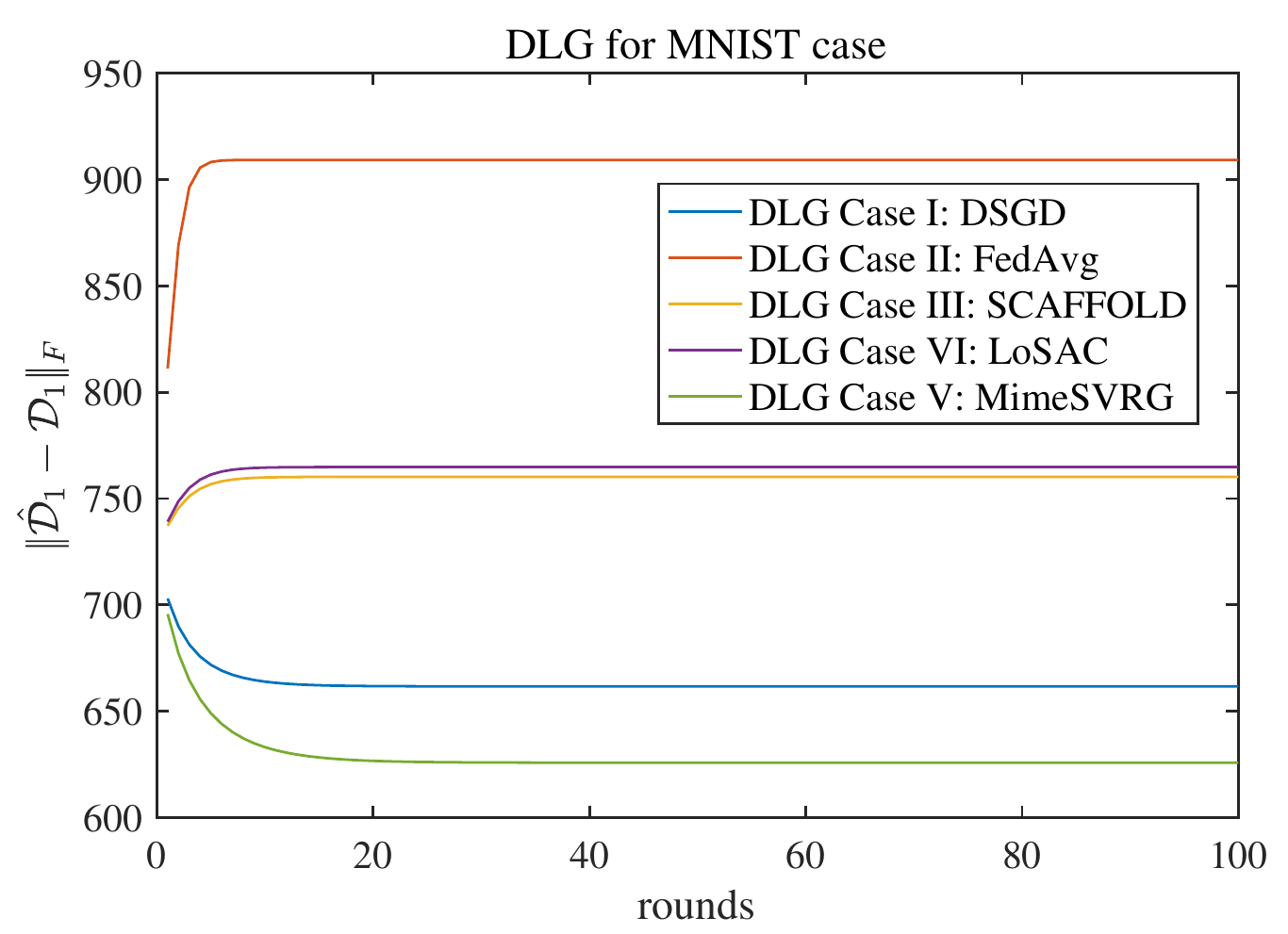}
			\end{subfigure}\\
		\end{minipage} \hfill 
		\begin{minipage}[t]{0.32\linewidth} 
			\centering
			\begin{subfigure}{1\textwidth}
				\centering
				\includegraphics[width=1.0\linewidth]{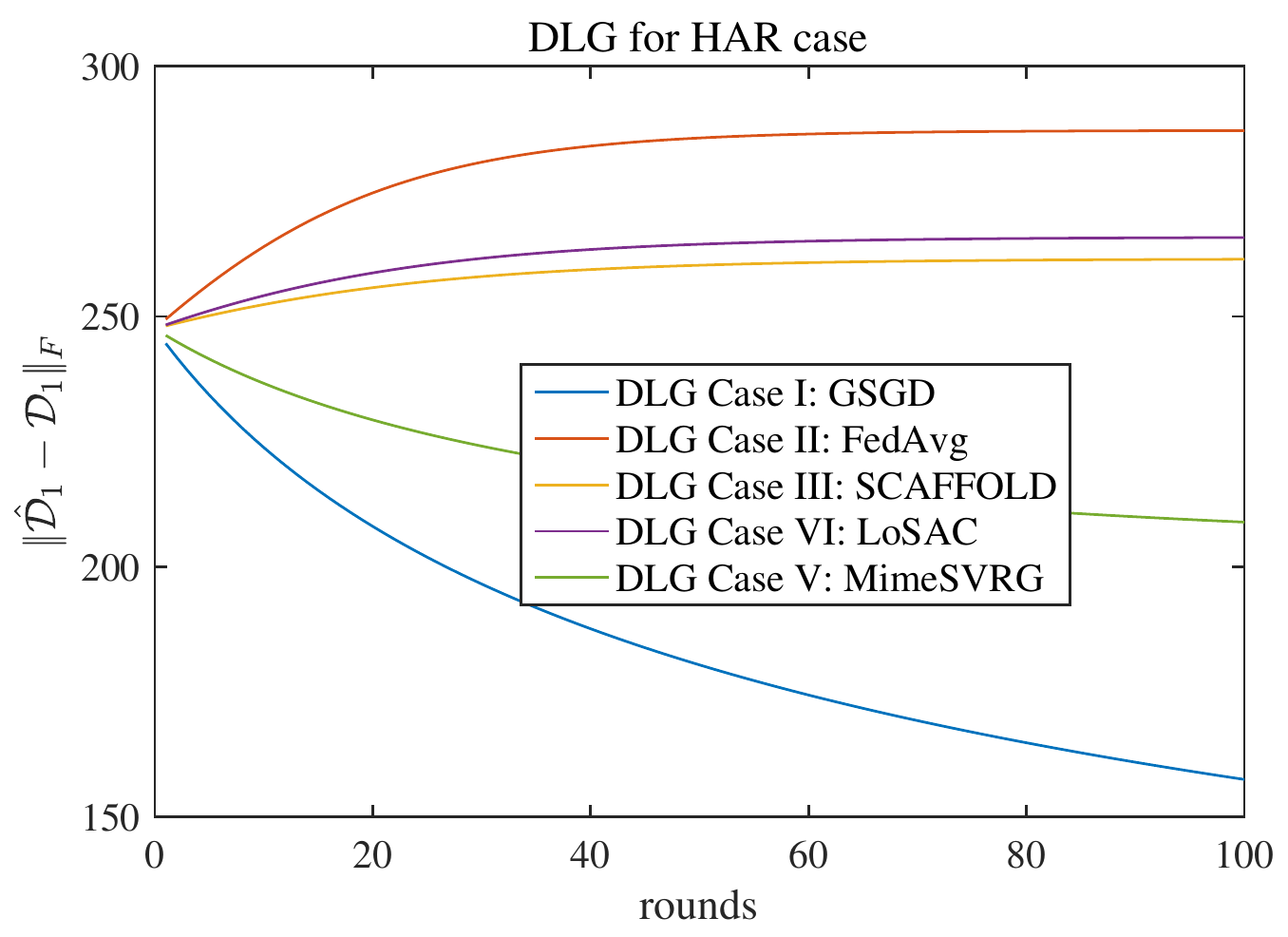}
			\end{subfigure}\\
		\end{minipage} \hfill			
		\begin{minipage}[t]{0.32\linewidth} 
			\centering
			\begin{subfigure}{1\textwidth}
				\centering
				\includegraphics[width=1.0\linewidth]{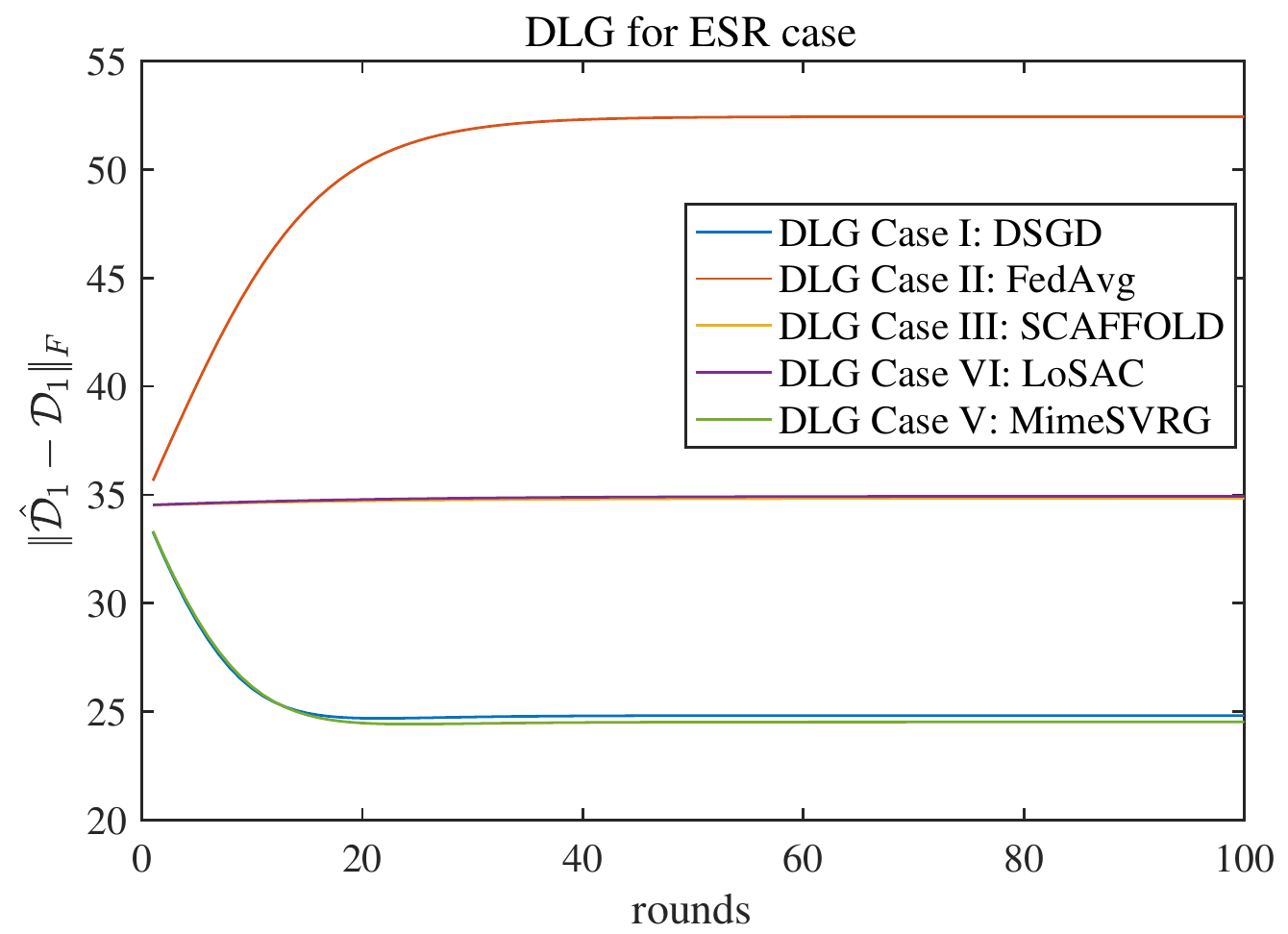}
			\end{subfigure}\\
		\end{minipage}%
		
		\caption{The performance evaluations on the defense against DLG. Binary logistic regression on the federated datasets is collaboratively performed over all clients. Moreover, GD is applied for solving the DLG problem to steal the dataset on the $1$st client. }
		\label{fig:dlg}
	\end{figure*}

	\begin{figure*}
		\centering
		\includegraphics[width=1.00\linewidth]{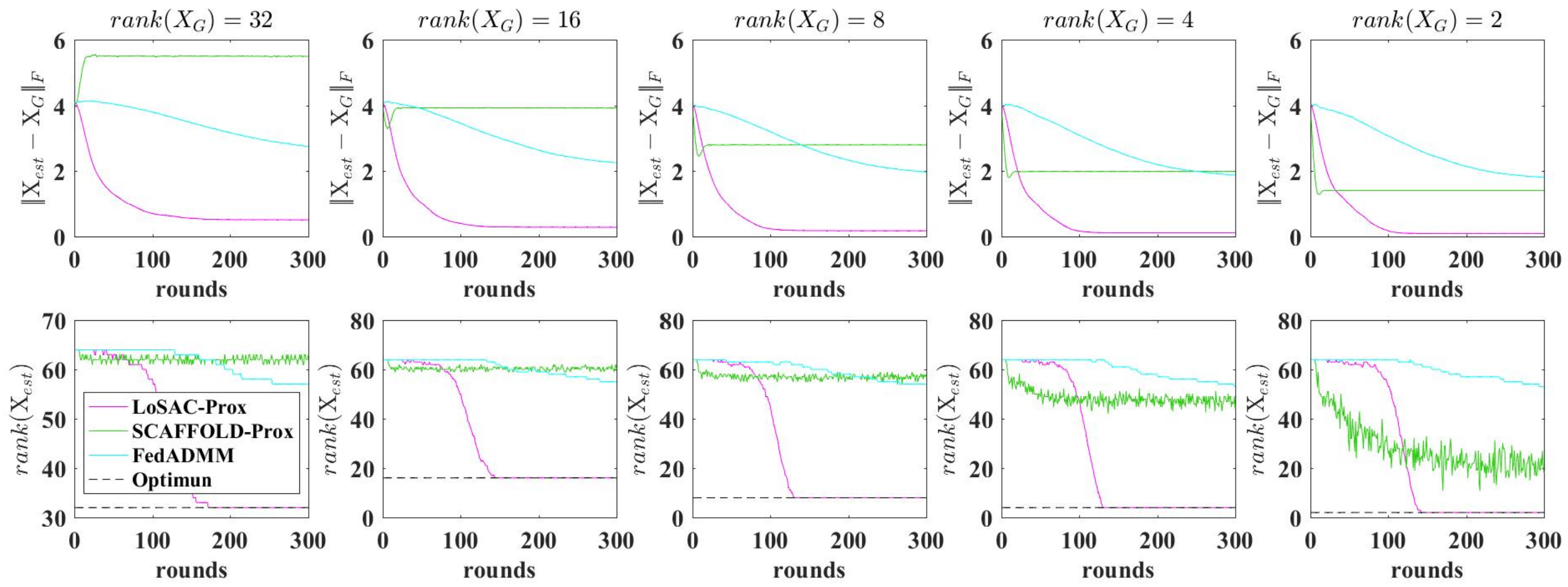}			
		\caption{The performance evaluations on the low rank matrix estimation problem.  Figures in the above row are the matrix recovery error   and the below figures are the rank of the recovery matrix.}
		\label{fig:nuclearnormprox}
	\end{figure*}

	\subsection{Low Rank Matrix Estimation}\label{sec:lrme}
	We further verify LoSAC for solving the problem of low rank matrix estimation via the comparisons with SCAFFOLD. First, the problem can be formulated in the following:
	\begin{equation}\label{lrme}
		\underset{X}{\text{min }} \nicefrac{1}{N} \sum_{i=1}^{N} f_i(X) + \lambda\norm{X}_*,
	\end{equation}
	where $f_i(X)=\sum_{j=1}^{n_i}(\langle X,D_j\rangle-y_j)^2$, $D_j\in\mathbb{R}^{d\times d}$ is the data sample, $y_j$ is the noisy observation and $\norm{\cdot}_*$ denotes the nuclear norm of a matrix to obtain a low rank matrix solution $X^*$. The problem  (\ref{lrme}) targets at recovering the low rank matrix $X\in\mathbb{R}^{d\times d}$ from the noisy observations $y_j$. The proximal version of the state-of-the-art algorithm SCAFFOLD, namely SCAFFOLD-Prox, and FedADMM are studied as the comparisons. In particular, FedADMM is known to be convenient and efficient to solve the nonsmooth optimization problem \cite{fedadmm,admm}. For LoSAC-Prox and SCAFFOLD-Prox, solving the above problem via the proximal operation, we can derive $X^*\leftarrow \text{ argmin}_Z \lambda\norm{Z}_*+\frac{1}{2\eta}\|Z-(X-\eta\tilde{\nabla }f_i(X))\|^2_F$, where we denote $\tilde{\nabla }f_i(X)$ as the global gradient estimate. Then, the above problem can be solved via $X^*=U\cdot\text{diag}\big\{\text{prox}_{\eta\norm{\cdot}_1}(\sigma)\big\}\cdot V^T$, where $U$, $\sigma$ and $V$ can be conveniently obtained via the singular value decomposition of the matrix $(X-\eta\tilde{\nabla }f_i(X))$. For FedADMM \cite{fedadmm} solving low
	rank matrix estimation, the details are in Appendix \ref{app:fedadmm}.

	We evaluate the algorithms on the synthetic dataset for simplicity with the known ground truth  $X_G\in\mathbb{R}^{d\times d}$, which is obtained as follows:
	\begin{equation}
		X_G = \begin{bmatrix}
			I_{rank}&0_{rank\times (d-rank)}  \\
			0_{(d-rank)\times rank}& 0_{(d-rank)\times(d-rank)}
		\end{bmatrix}				
	\end{equation}		  
	As for the synthetic dataset generation, each element of a data sample $D_j$ is randomly  generated by $\mathcal{N}(0.1,1)$. Moreover, $y_j\in\mathbb{R}$ is generated by $y_j=\langle X_G,D_j\rangle+\mathcal{N}(0,0.1)$. Specifically, we set $d=64$ for $X_G$ and five scenarios of the matrix rank are studied, i.e., $rank(X_G)=[32,16,8,4,2]$.  We set $(N,S)=(100,10)$ and the local dataset division to be $M=5$ for each algorithm. We tune the step size $\eta=2\times10^{-3}$ for each algorithm to yield as the best performance as possible.
	We adopt two metrics for performance evaluation, namely the recovery matrix error $\norm{X_{est}-X_G}_F$ and the rank of the recovery matrix $rank(X_{est})$ (we count the number of singular values that are greater than $10^{-3}$). 
	
	The results are shown in Fig. \ref{fig:nuclearnormprox}. It can be seen although ADMM has shown to conveniently and
	efficiently solve the nonsmooth optimization problem, with FedOpt settings (e.g., partial client
	participation), its performance is substantially degraded. In particular for SCAFFOLD, although
	it performs satisfactory results in the overall performance and ablation study, SCAFFOLD works
	poor in the low rank matrix estimation and cannot precisely recover the ground truth matrix, i.e.,
	while the recovery matrix error is large, the rank of the estimated matrix significantly differs from
	the rank of the ground truth matrix.  For LoSAC-Prox, it has shown the superiority in solving the problem. The recovery matrix $X_{est}$ matches very well with the ground truth matrix  $X_G$, i.e., it has the smallest recovery error and the \textit{EXACT} rank with $X_G$.

	\section{Conclusion}
	Due to the  data heterogeneity and client sampling, the performance of  federated optimization suffers from degradation. {Although  there are several works attempting to mitigate these problems, none of them could well address them. We proposed a new FedOpt algorithm LoSAC  for handling the challenges by compactly estimating the global gradient. Moreover, we extend LoSAC to its proximal version for solving a wider class of problems.} We demonstrate the effectiveness of LoSAC via theoretical guarantees and the empirical studies. It shows that LoSAC equips with the strong ability in handling the model divergence problem, and the high communication and computation efficiency over the state-of-the-art methods. Especially in the low-rank matrix estimation problem, LoSAC has demonstrated its superior performances over SCAFFOLD and FedADMM, i.e., it can recover very well the true matrix with the exact rank.  It is worth mentioning that LoSAC  has the  defense ability against the information leakage from the gradient.


\bibliographystyle{ACM-Reference-Format}
\bibliography{iclr2021_conference}

\appendix

\section{Useful Lemmas}
For proving Lemma \ref{lemma4}, we provide the following useful lemmas. First, we present the Lemma \ref{lemma1}, of which the relaxed triangle inequalities have provided quite important tools for evaluating the update progress in each round. Based on Lemma \ref{lemma1}, the inequalities in Lemma \ref{lemma2}  can be derived.

\begin{lemma}\label{lemma1}
	For vectors $\{a_1,\dots,a_m\}$ in $\mathbb{R}^d$, the following inequalities hold:
	\begin{equation}\label{iq1}
		\norm{a_i+a_j}^2\leq (1+\alpha)\norm{a_i}^2+(1+\frac{1}{\alpha})\norm{a_j}^2, \text{ and}
	\end{equation}
	\begin{equation}\label{iq2}
		\left\|\sum\nolimits_{i=1}^{m}a_i\right\|^2\leq m\sum\nolimits_{i=1}^{m}\norm{a_i}^2,
	\end{equation}
	where $\alpha>0$.
\end{lemma}

\begin{lemma}\label{lemma2}
	Suppose $f$ is a strongly convex function with $\mu>0$ and has $L$-smoothness Lispschitz gradient, then the following inequalities holds:
	\begin{equation}\label{ie1}
		\begin{aligned}
			\left\langle \nabla f(x),z-y \right\rangle\leq& f(z)-f(y)+(L+\beta L-\nicefrac{\mu}{4})\norm{y-z}^2+(1+\beta^{-1})L\norm{z-x}^2,
		\end{aligned}
	\end{equation}
	and moreover 
	\begin{equation}\label{ie2}
		\begin{aligned}
			\left\langle \nabla f(x),z-y \right\rangle\geq& f(z)-f(y)+\nicefrac{\mu}{[2(1+\alpha)]}\norm{y-z}^2-(\nicefrac{L}{2}+\nicefrac{\mu}{2\alpha})\norm{z-x}^2.
		\end{aligned}
	\end{equation}
	where $\beta>0$ and $\alpha>0$.
\end{lemma}

\textit{Proof}: Lemma \ref{lemma2} can be derived with Lemma \ref{lemma1} and the definitions of $L$-smoothness and $\mu$-convexity. Specifically, since $f$ is strongly convex and has $L$-smooth gradient, it implies that $L\geq \mu$, which follows:
\begin{equation}
	\begin{aligned}
		\langle \nabla& f(x),x-y \rangle\leq f(x)-f(y)+\nicefrac{L}{2}\norm{x-y}^2\text{ and }\\
		& \langle \nabla f(x),z-x \rangle\leq f(z)-f(x)-\nicefrac{\mu}{2}\norm{z-x}^2
	\end{aligned}
\end{equation}
hold by the definitions of $L$-smoothness and strongly convexity respectively. Combining the two inequalities yields:
\begin{equation}\label{ie3}
	\begin{aligned}
		\langle \nabla f(x),z-y \rangle\leq f(z)-f(y)+L\norm{x-y}^2-\frac{\mu}{4}\norm{z-y}^2,
	\end{aligned}
\end{equation}
where we have used the triangle inequality. We further use the triangle inequality as follows:
\begin{equation}
	\norm{x-y}^2\leq (1+\beta)\norm{y-z}^2+(1+\beta^{-1})\norm{z-x}^2,
\end{equation}
which we substitute into (\ref{ie3}) that can lead to the desired result in (\ref{ie1}). For the second inequality in (\ref{ie2}), it follows from the first inequality in (\ref{ie3}) that
\begin{equation}\label{ie4}
	\langle \nabla f(x),z-y \rangle\geq f(z)-f(y)-\frac{L}{2}\norm{x-z}^2+\frac{\mu}{2}\norm{y-x}^2.
\end{equation}
Furthermore, the triangle inequality with $\alpha>0$ is utilized and it yields:
\begin{equation}\label{ie5}
	\norm{y-x}^2\geq\nicefrac{1}{(1+\alpha)}\norm{y-z}^2-\nicefrac{1}{\alpha}\norm{z-x}^2.
\end{equation}
By combining (\ref{ie4}) and (\ref{ie5}), we can obtain the desired result in (\ref{ie2}) and this completes the proof of Lemma \ref{lemma2}.

%

\section{Proof of Lemma \ref{lemma4}}\label{eachroundprogress}

\begin{proof}
	The proof of Lemma \ref{lemma4} is based on the results of Lemmas \ref{lemma1} and \ref{lemma2}, which are in the Appendix A. We first recall the following inequalities hold respectively:
	\begin{equation}\label{Lsmooth}
		f(y)\leq f(x)+\left\langle \nabla f(x),y-x \right\rangle+\frac{L}{2}\norm{x-y}^2,\text{ and}
	\end{equation}
	\begin{equation}\label{ie6}
		\norm{\nabla f(y)-\nabla f(x)}\geq\mu\norm{y-x}.
	\end{equation}

	Then we start with evaluating $\Delta x^r_i$ within $T$ iterations given as
	\begin{equation}
		\small
		\Delta x^r_i = -\eta\sum\nolimits_{t=0}^{T-1}\{\frac{1}{N} \sum\nolimits_{n=1}^{N}\nabla f_n(\{z^t_n\}) -g_{i,j_t}(z^t_{i,j_t})+g_{i,j_t}(x^t_i)\}.
	\end{equation}
	Next, we aim to evaluate the progress in one round. To be specific:
	\begin{equation}\label{base}
		\begin{aligned}
			\mathbb{E}\norm{x^{r+1}-x^*}^2&=\mathbb{E}\norm{x^{r+1}-x^r}^2+\mathbb{E}\norm{x^{r}-x^*}^2+2\mathbb{E}\langle x^{r+1}-x^r,x^r-x^* \rangle.\\
		\end{aligned}
	\end{equation}
	We first expand the term $\mathbb{E}\norm{x^{r+1}-x^r}^2$ as follows, then the upper bound can be further derived (with $c_0:=\nicefrac{\eta^2S^2T}{N^3}$):
	\begin{equation}\label{ie7}
		\begin{aligned}
			&\mathbb{E}\norm{x^{r+1}-x^r}^2\leq\frac{S^2}{N^3}\sum_{i=1}^{N}\mathbb{E}\norm{\Delta x^r_i}^2\leq c_0\sum_{i,t}\mathbb{E}\|\frac{1}{N} \sum_{n=1}^{N}\nabla f_n(\{z^t_n\}) -g_{i,j_t}(z^t_{i,j_t})+g_{i,j_t}(x^t_i)\|^2\\
			&c_0\sum_{i,t}\mathbb{E}\|\frac{1}{N} \sum_{n=1}^{N}\nabla f_n(\{z^t_n\})-\frac{1}{N} \sum_{n=1}^{N}\nabla f_n(x^*)-g_{i,j_t}(z^t_{i,j_t})+g_{i,j_t}(x^*)+g_{i,j_t}(x^t_i)-g_{i,j_t}(x^*)\|^2,
		\end{aligned}
	\end{equation}
	thus by defining $c_1:=\nicefrac{3\eta^2S^2Tn_m}{N^3}$ we have
	\begin{equation}
		\begin{aligned}
			&\mathbb{E}\norm{x^{r+1}-x^*}^2\leq\frac{c_1}{N}\sum_{i,t,n,j}\mathbb{E}\|\nabla f_{n,j}(z^{i,t}_{n,j})-\nabla f_{n,j}(x^*)\|^2\\
			&+c_1n_m\sum_{i,t}\mathbb{E}\|\nabla f_{i,j_t}(z^t_{i,j_t})-\nabla f_{i,j_t}(x^*)\|^2+c_1n_m\sum_{i,t}\mathbb{E}\|\nabla f_{i,j_t}(x^t_{i})-\nabla f_{i,j_t}(x^*)\|^2,
		\end{aligned}
	\end{equation}
	where we have adopted the relaxed triangle inequality in Lemma \ref{lemma1}. By using the condition of Lipschitz continuity, it yields:
	\begin{equation}
		\begin{aligned}
			\mathbb{E}&\norm{x^{r+1}-x^r}^2\leq\frac{c_1L}{N}\sum_{i,t,n,j}\mathbb{E}\|z^{i,t}_{n,j}-x^*\|^2+c_1n_mL\sum_{i,t}\mathbb{E}\|z^t_{i,j_t}-x^*\|^2+c_1n_mL\sum_{i,t}\mathbb{E}\|x^t_{i}-x^*\|^2.
		\end{aligned}
	\end{equation}
	The above equality has used the fact that $x^*$ is the optimal point which satisfies the first-order condition, i.e., $\nicefrac{1}{N} \sum_{n=1}^{N}\nabla f_n(x^*)=0$. Moreover, the above inequalities include the terms $\mathbb{E}\|z^{i,t}_{n,j}-x^*\|^2$ and $\mathbb{E}\|z^{i,t}_{n,j}-x^*\|^2$, directly evaluating them is extremely difficult since they are randomly selected and updated. Note that both the gradient variance and second-order moment are bounded, it implies from (\ref{ie6}) that:
	\begin{equation}\label{ie8}
		\norm{y-x}\leq(\delta_f-\sigma_f)\cdot\nicefrac{2}{\mu}:=\mathcal{M}_v
	\end{equation}
	for any $x$ and $y$ in the domain of $f$. Therefore, we use (\ref{ie8}) for bounding $\mathbb{E}\|z^{i,t}_{n,j}-x^*\|^2$ and $\mathbb{E}\|z^{i,t}_{n,j}-x^*\|^2$. Moreover, to bound the last term $\mathbb{E}\|x^t_{i}-x^*\|^2$, the one iteration progress  for the client $i$ can be expanded as follows:
	\begin{equation}
		\begin{aligned}
			\mathbb{E}\|x^{t+1}_{i}-x^*\|^2&\leq\mathbb{E}\|x^{t+1}_{i}-x^{t}_i\|^2+\mathbb{E}\|x^t_{i}-x^*\|^2+2\mathbb{E}\langle x^{t+1}_i-x^t_i,x^t_i-x^*\rangle,
		\end{aligned}
	\end{equation}
	which has the same form with (\ref{base}). We omit the duplicated procedure and use the inequality $\mathbb{E}\|x^t_{i}-x^*\|^2\leq\mathcal{M}_v$. Hence, by combining (\ref{ie7}) and (\ref{ie8}), the following result can be derived:
	\begin{equation}\label{ie9}
		\mathbb{E}\norm{x^{r+1}-x^r}^2\leq\frac{3\eta^2S^2T^2n_mL\mathcal{M}_v(|\mathcal{D}|+2n_mN)}{N^3}.
	\end{equation}
	
	We next evaluate the term $2\mathbb{E}\langle x^{r+1}-x^r,x^r-x^* \rangle$, by expanding it and defining $c_2:=\nicefrac{2\eta S}{N^2}$, it leads to
	\begin{equation}\label{ie10}
		\begin{aligned}
			  2\mathbb{E}\langle x^{r+1}-x^r,&x^r-x^* \rangle=\frac{-c_2}{N}\sum_{i,t,n,j}[\mathbb{E}\langle \nabla f_{n,j}(z^{i,t}_{n,j}),x^r-x^* \rangle]\\ &-c_2\sum_{i,t}\mathbb{E}\langle \nabla f_{i}(x^{t}_{i}),x^r-x^* \rangle+c_2\sum_{i,t,j}\mathbb{E}\langle \nabla f_{i,j}(z^{t}_{i,j}),x^r-x^* \rangle.	
		\end{aligned}
	\end{equation}
	Then, we evaluate (\ref{ie10}) term by term  for convenience. For the first and third terms, they can bounded by applying (\ref{ie2}), while the second term can be bounded by utilizing (\ref{ie1}). Therefore, it yields the inequalities term by term in the following:
	
	The evaluation of the first term ($c_3:=\nicefrac{\eta ST}{N}$):
	\begin{equation}\label{ie11}
		\begin{aligned}
			\frac{-2\eta S}{N^3}\sum_{i,t,n,j}\mathbb{E}\langle \nabla f_{n,j}(z^{i,t}_{n,j}),x^r-x^* \rangle\leq-2&c_3\mathbb{E}\{f(x^r)-f(x^*)\}-\frac{c_3|\mathcal{D}|\mu}{N(1+\alpha)}\mathbb{E}\|x^r-x^*\|^2\\
			&+\left\{L+\frac{\mu}{\alpha} \right\}\frac{c_3}{N^2T}\sum_{i,t,n,j}\mathbb{E}\|z^{i,t}_{n,j}-x^r\|^2;
		\end{aligned}
	\end{equation} 
	The evaluation of the second term:
	\begin{equation}\label{ie12}
		\begin{aligned}
			\frac{2\eta S}{N^2}\sum_{i,t,j}\mathbb{E}\langle \nabla f_{i,j}(z^{t}_{i,j}),x^r-x^* \rangle\leq &2c_2\mathbb{E}\{f(x^r)-f(x^*)\}+\frac{2c_3|\mathcal{D}|}{N}(L+\beta L-\frac{\mu}{4})\mathbb{E}\|x^r-x^*\|^2\\
			&+\frac{2c_3(1+\beta^{-1})}{N}\sum_{i,t,j}\mathbb{E}\|z^t_{i,j}-x^r\|^2.
		\end{aligned}
	\end{equation}
	The evaluation of the third term:
	\begin{equation}\label{ie13}
		\begin{aligned}
			\frac{-2\eta S}{N^2}\sum_{i,t}\mathbb{E}\langle \nabla f_{i}(x^{t}_{i}),x^r-x^* \rangle&\leq-2c_3\mathbb{E}\{f(x^r)-f(x^*)\}-\frac{c_3\mu}{1+\alpha}\mathbb{E}\|x^r-x^*\|^2\\
			&+\left\{L+\frac{\mu}{\alpha} \right\}\frac{c_3}{NT}\cdot\sum_{i,t}\mathbb{E}\|x^t_i-x^r\|^2.
		\end{aligned}
	\end{equation}
	
	Consequently, if $T$ is sufficiently large, we assume all the locally stored variables $\{z_{i,j}\}$ in client $i$ are expected to be participated for the update at least once, and we assume the update times is $\tau$ for all locally stored parameters $\{z_{i,j}\}$. Moreover, we denote $t'$ as the latest update iteration for evaluating $\mathbb{E}\|z^{i,t}_{n,j}-x^*\|^2$, $t''$ as the second latest update, and so on. First of all, the inequalities (\ref{ie11})-(\ref{ie13}) can be rewritten with the relaxed triangle inequality and listed as follows:
	
	The evaluation of the first term ($c_4:=\nicefrac{\eta ST}{N}$):
	\begin{equation}\label{ie17}
		\begin{aligned}
			\frac{-2\eta S}{N^3}\sum_{i,t,n,j}&\mathbb{E}\langle \nabla f_{n,j}(z^{i,t}_{n,j}),x^r-x^* \rangle\leq-2c_4\mathbb{E}\{f(x^r)-f(x^*)\}-\frac{c_4|\mathcal{D}|\mu}{N(1+\alpha)}\mathbb{E}\|x^r-x^*\|^2\\
			&+\left(L+\frac{\mu}{\alpha} \right)\frac{(1+\beta)c_4|\mathcal{D}|}{N}\mathbb{E}\|x^r-x^*\|^2+\left(L+\frac{\mu}{\alpha} \right)\frac{(1+\beta^{-1})c_4}{N^2T}\sum_{i,t,n,j}\mathbb{E}\|z^{i,t}_{n,j}-x^*\|^2;
		\end{aligned}
	\end{equation} 
	For the evaluation of the second term, it yields in the sequel:
	\begin{equation}\label{ie18}
		\begin{aligned}
			\frac{2\eta S}{N^2}\sum_{i,t,j}\mathbb{E}\langle \nabla f_{i,j}(z^{t}_{i,j}),x^r-x^* \rangle\leq2c_4&\mathbb{E}\{f(x^r)-f(x^*)\}-\frac{c_4|\mathcal{D}|\mu}{N(1+\alpha)}\mathbb{E}\|x^r-x^*\|^2\\
			&+\frac{c_4}{NT}\left(L+\frac{\mu}{\alpha}\right)\sum_{i,t,j}\mathbb{E}\|z^t_{i,j}-x^*\|^2.
		\end{aligned}
	\end{equation}
	The evaluation of the third term:
	\begin{equation}\label{ie19}
		\begin{aligned}
			\frac{-2\eta S}{N^2}&\sum_{i,t}\mathbb{E}\langle \nabla f_{i}(x^{t}_{i}),x^r-x^* \rangle\leq-2c_4\mathbb{E}\{f(x^r)-f(x^*)\}-\frac{c_4\mu}{1+\alpha}\mathbb{E}\|x^r-x^*\|^2\\
			&+\left(L+\frac{\mu}{\alpha} \right)(1+\beta)c_4\mathbb{E}\|x^r-x^*\|^2+\left(L+\frac{\mu}{\alpha} \right)\frac{(1+\beta^{-1})c_4}{NT}\sum_{i,t}\mathbb{E}\|x^t_i-x^*\|^2.
		\end{aligned}
	\end{equation}
	
	We first evaluate the one iteration progress for the term $\mathbb{E}\|z^{i,t}_{n,j}-x^*\|^2$ in (\ref{ie17}), which can be expanded as follows:
	\begin{equation}\label{ie20}
		\begin{aligned}
			\mathbb{E}&\|z^{i,t'}_{n,j}-x^*\|^2=\mathbb{E}\|z^{i,t'}_{n,j}-z^{i,t''}_{n,j}\|^2+2\mathbb{E}\langle z^{i,t'}_{n,j}-z^{i,t''}_{n,j},z^{i,t''}_{n,j}-x^*\rangle+\mathbb{E}\|z^{i,t''}_{n,j}-x^*\|^2.
		\end{aligned}
	\end{equation}
	Hence, it can be evaluated term by term for convenience. Note $z^{i,t'}_{n,j}$ is updated based on $z^{i,t''}_{n,j}$ \footnote{To be more exact, $j$ should be $j_{t'}$, for notational simplicity, we use $j$ since it is nontrivial in our derivation.}, namely,
	\begin{equation}
		z^{i,t'}_{n,j}=z^{i,t''}_{n,j}-\eta\{\frac{1}{N}\sum_{m=1}^{N}\nabla f_m(\{z'''_m\})-g_{i,j}(z^{i,t'''}_{n,j})+g_{i,j}(z^{i,t''}_{n,j})\}.
	\end{equation}
	Furthermore,   $\mathbb{E}\|z^{i,t'}_{n,j}-z^{i,t''}_{n,j}\|^2$ can be bounded as follows:
	\begin{equation}
		\begin{aligned}
			\mathbb{E}&\|z^{i,t'}_{n,j}-z^{i,t''}_{n,j}\|^2\leq3\eta^2\mathbb{E}\|\frac{1}{N}\sum_{m=1}^{N}\nabla f_m(\{z^{t'''}_m\})-\frac{1}{N}\sum_{m=1}^{N}\nabla f_m(x^*)\|^2\\
			&\quad\quad+3\eta^2\mathbb{E}\|g_{i,j}(z^{i,t'''}_{n,j})-g_{i,j}(x^*)\|^2+3\eta^2\mathbb{E}\|g_{i,j}(z^{i,t''}_{n,j})-g_{i,j}(x^*)\|^2\\
			&\leq \frac{3\eta^2n_mL}{N}\sum_{m,j}\mathbb{E}\|z^{i,t'''}_{n,j}-x^*\|^2+3\eta^2n^2_mL\mathbb{E}[\|z^{i,t'''}_{n,j}-x^*\|^2+\|z^{i,t''}_{n,j}-x^*\|^2],
		\end{aligned}
	\end{equation}
	where the first inequality has used the relaxed triangle inequality. Moreover, recall $\|y-x\|^2\leq\mathcal{M}_v$ for all $x$ and $y$ in the domain of $f_{i,j}$. Hence, we have 
	\begin{equation}
		\mathbb{E}\|z^{i,t'}_{n,j}-z^{i,t''}_{n,j}\|^2\leq9\eta^2n^2_mL\mathcal{M}_v.
	\end{equation}
	Next, we evaluate the second term $2\mathbb{E}\langle z^{i,t'}_{n,j}-z^{i,t''}_{n,j},z^{i,t''}_{n,j}-x^*\rangle$, it can be derived in the following:
	\begin{equation}
		\begin{aligned}
			2\mathbb{E}\langle z^{i,t'}_{n,j}&-z^{i,t''}_{n,j},z^{i,t''}_{n,j}-x^*\rangle=\frac{-2\eta}{N}\cdot\sum_{m,p}\mathbb{E}\langle\nabla f_{m,p}(z^{i,t'''}_{m,p}),z^{i,t''}_{n,j}-x^*\rangle+2\eta n_i\cdot\\
			&\mathbb{E}\langle\nabla f_{i,j}(z^{i,t'''}_{i,j}),z^{i,t''}_{n,j}-x^*\rangle-2\eta n_i\mathbb{E}\langle\nabla f_{i,j}(z^{i,t''}_{i,j}),z^{i,t''}_{n,j}-x^*\rangle.
		\end{aligned}
	\end{equation}
	Therefore, we can use the inequalities in Lemma \ref{lemma2}  term by term for further evaluations:
	
	\underline{\textit{The first term:}} 
	
	\begin{equation}
		\begin{aligned}
			\frac{-2\eta}{N}\sum_{m,p}\mathbb{E}\langle\nabla f_{m,p}(z^{i,t'''}_{m,p}),z^{i,t''}_{n,j}-x^*\rangle&\leq\frac{-2\eta}{N}\cdot\mathbb{E}\left(f(\{z^{t''}_m\})-f(x^*)\right)-\frac{\mu\eta|\mathcal{D}|}{N(1+\alpha)}\mathbb{E}\|z^{i,t''}_{n,j}-x^*\|^2\\
			&+\frac{\eta}{N}\left(L+\frac{\mu}{\alpha}\right)\sum_{m,p}\mathbb{E}\|z^{i,t''}_{n,j}-z^{i,t'''}_{n,j}\|^2;
		\end{aligned}
	\end{equation}
	
	\underline{\textit{The second term:}} 
	
	\begin{equation}
		\begin{aligned}
			2\eta n_i\langle\nabla f_{i,j}(z^{i,t'''}_{i,j}),z^{i,t''}_{n,j}-x^*\rangle&\leq2\eta n_i\mathbb{E}\left(f_{i,j}(z^{i,t''}_{i,j})-f_{i,j}(x^*) \right)+2\eta n_i\left(L+\beta L-\frac{\mu}{4}\right)\mathbb{E}\|z^{i,t''}_{i,j}-x^*\|^2\\
			&+2\eta n_i(1+\beta^{-1})L\mathbb{E}\|z^{i,t'''}_{i,j}-z^{i,t''}_{i,j}\|^2;
		\end{aligned}
	\end{equation}
	
	\underline{\textit{The third term:}} 
	
	\begin{equation}
		\begin{aligned}
			-2\eta n_i\langle\nabla f_{i,j}(z^{i,t''}_{i,j}),z^{i,t''}_{n,j}-x^*\rangle&\leq-2\eta n_i\cdot\mathbb{E}\left(f_{i,j}(z^{i,t''}_{i,j})-f_{i,j}(x^*) \right)-\frac{\eta n_i}{1+\alpha}\mathbb{E}\|z^{i,t''}_{i,j}-x^*\|^2\\
			&+\eta n_i\left(L+\frac{\mu}{\alpha}\right)\mathbb{E}\|z^{i,t''}_{n,j}-z^{i,t''}_{i,j}\|^2.
		\end{aligned}
	\end{equation}
	Substituting the above results into (\ref{ie20}), we can obtain the following:
	\begin{equation}
		\begin{aligned}
			\mathbb{E}\|z^{i,t'}_{n,j}-x^*\|^2\leq&(1-\eta\nu)\mathbb{E}\|z^{i,t''}_{n,j}-x^*\|^2+\eta\lambda+9\eta^2n^2_mL\mathcal{M}_v,
		\end{aligned}
	\end{equation}
	where $\nu$ and $\lambda$ are provided respectively as
	\begin{equation}
		\nu=\frac{\mu(|\mathcal{D}|+Nn_{m})}{N(1+\alpha)}-2 n_{min}\left(L+\beta L-\frac{\mu}{4}\right) \text{ and}
	\end{equation}
	\begin{equation}
		\lambda=\left(n_{m}+\frac{|\mathcal{D}|}{N}\right)\left(L+\frac{\mu}{\alpha}\right)\mathcal{M}_v+2n_{m}(1+\beta^{-1})L\mathcal{M}_v.
	\end{equation}
	Recall that we have assumed the update time for $z$ is $\tau$, hence we further have the following inequality
	\begin{equation}\label{ie21}
		\begin{aligned}
			\mathbb{E}\|z^{i,t'}_{n,j}-x^*\|^2&\leq(1-\eta\nu)^{\tau}\left(\mathcal{M}_v+\frac{\lambda+9\eta n^2_mL\mathcal{M}_v}{v}\right)-\frac{\lambda+9\eta n^2_mL\mathcal{M}_v}{v}\\
			&\leq \eta(\tau \nu\mathcal{M}_v+\tau \lambda)+9\eta^2 n^2_mL\mathcal{M}_v
		\end{aligned}
	\end{equation}
	holds when $\frac{1}{2}\leq\eta\nu< 1$. Using the same strategy for the derivation, we have the following bounds for $\mathbb{E}\|z^t_{i,j}-x^*\|^2$ and $\mathbb{E}\|x^t_i-x^*\|^2$:
	\begin{equation}\label{ie22}
		\mathbb{E}\|z^t_{i,j}-x^*\|^2\leq\eta(\tau \nu\mathcal{M}_v+\tau \lambda)+9\eta^2 n^2_mL\mathcal{M}_v \text{ and}
	\end{equation}
	\begin{equation}\label{ie23}
		\mathbb{E}\|x^t_i-x^*\|^2\leq\eta(T \nu\mathcal{M}_v+T \lambda)+9\eta^2 n^2_mL\mathcal{M}_v, \text{ respectively.}
	\end{equation}
	With inequalities (\ref{ie17}), (\ref{ie18}), (\ref{ie19}), (\ref{ie20}), (\ref{ie21}), (\ref{ie22}) substituting into (\ref{base}), the following one round progress can be derived:
	\begin{equation}
		\begin{aligned}
			&\mathbb{E}\|x^{r+1}-x^*\|^2\leq\frac{-2\eta ST}{N}\mathbb{E}\{f(x^r)-f(x^*)\}+(1-\eta h)\mathbb{E}\|x^{r}-x^*\|^2+\lambda'\eta^2+\nu'\eta^3,
		\end{aligned}
	\end{equation}
	where we have defined $h$, $\lambda'$ and $\nu'$ respectively with $T\geq\tau$ as follows:
	\begin{equation}
		h=\frac{\mu ST(2|\mathcal{D}|+N)}{N^2(1+\alpha)}-\left(L+\frac{\mu}{\alpha}\right)\frac{ST(1+\beta)(|\mathcal{D}|+N)}{N^2},
	\end{equation}
	\begin{equation}
		\begin{aligned}
			\lambda'=(&T \nu_2\mathcal{M}_v+T \lambda)\left(L+\frac{\mu}{\alpha}\right)\cdot\nicefrac{[(\beta^{-1}+2)ST|\mathcal{D}|+(\beta^{-1}+1)STN]}{N^2},
		\end{aligned} 
	\end{equation}
	\begin{equation}
		\nu'=9n^2_mL\mathcal{M}_v\left(L+\frac{\mu}{\alpha}\right)\frac{(\beta^{-1}+2)ST|\mathcal{D}|+(\beta^{-1}+1)STN}{N^2}.
	\end{equation}
	
\end{proof}

\textbf{Discussion: } For $h$, which can be rewritten as follows:
\begin{equation}
	h=\frac{ST}{N^2}\left\{\frac{\mu (2|\mathcal{D}|+N)}{(1+\alpha)}-\left(L+\frac{\mu}{\alpha}\right){(1+\beta)(|\mathcal{D}|+N)}\right\},
\end{equation}
hence, to let $h_2\rightarrow0$, the following must be satisfied
\begin{equation}\label{ie25}
	\frac{\mu}{L}\geq \frac{(1+\beta)(|\mathcal{D}|+N)}{\nicefrac{2|\mathcal{D}|}{(1+\alpha)}-\nicefrac{(1+\beta)(|\mathcal{D}|+N)}{\alpha}}.
\end{equation} 
In fact, if $\alpha>\frac{1+\beta}{1-\beta}$, (\ref{ie25}) may be satisfied. For the convergence study, we assume all these conditions are satisfied for simplicity.

\section{Proof of Theorem \ref{theorem1}}\label{convergenceproof}
\begin{proof}
	according the inequality in Lemma \ref{lemma4}, we have the following, 
	\begin{equation}
		\begin{aligned}
			p_{r+1}\leq\frac{-2\eta ST}{N}&\Phi_r+(1-\eta h)p_{r}+\lambda'\eta^2+\nu'\eta^3,
		\end{aligned}
	\end{equation}
	where we define $\Phi_r:=\mathbb{E}\{f(x^r)-f(x^*)\}$.	Hence, we  construct a positive sequence $\{w_r\}$ defined by $w_r=(1-\eta h_2)^{-r}$, and it follows:
	\begin{equation}\label{ie28}
		\frac{1}{W_R}\sum\nolimits_{r=0}^{R}w_r\Phi_r\leq\frac{Np_0}{2ST\eta W_R}+\frac{N\lambda'\eta}{2ST},
	\end{equation}
	where we have defined $W_R:=\sum\nolimits_{r=0}^{R}w_r$. Note when $R\geq\nicefrac{1}{3\eta h}$, $(1-\eta h)^{R+1}\leq\text{exp}\{-\eta hR\}\leq\text{exp}\{-\nicefrac{1}{3}\}\leq\nicefrac{3}{4}$, it leads to $W_R=\nicefrac{1}{4\eta h}(1-\eta h)^{-R}$. Substituting the  results into (\ref{ie28}), the weighted one round progress can be further simplifies as follows:
	\begin{equation}
		\begin{aligned}
			\frac{1}{W_R}\sum\nolimits_{r=0}^{R}w_r\Phi_r\leq\frac{2hNp_0}{ST}\text{exp}\{-\eta hR\}+\frac{N\lambda'\eta}{2ST}.
		\end{aligned}
	\end{equation}
	Therefore, we discuss the following two cases for the choice of the step size $\eta$:
	
	\begin{itemize}[leftmargin=*]
		\item if $\nicefrac{1}{3hR}\leq\tilde{\eta}\leq\nicefrac{\text{log}\{\text{max}(1,4h^2p_0/\lambda)\}}{hR}$, then we select the step size to be $\eta=\tilde{\eta}$ and it follows
		\begin{equation}
			\begin{aligned}
				\frac{1}{W_R}\sum\nolimits_{r=0}^{R}w_r\Phi_r\leq\frac{2hNp_0}{ST}\text{exp}\{-\tilde{\eta} hR\}+\mathcal{O}\left(\frac{N\lambda'}{2SThR} \right),
			\end{aligned}
		\end{equation}
		\item if $\tilde{\eta}>\nicefrac{\text{log}\{\text{max}(1,4h^2p_0/\lambda)\}}{hR}$, then we select the step size to be $\eta=\nicefrac{\text{log}\{\text{max}(1,4h^2p_0/\lambda)\}}{hR}$, and $\nicefrac{1}{W_R}\sum\nolimits_{r=0}^{R}w_r\Phi_r\leq\mathcal{O}(\frac{N\lambda'}{2SThR} ).$
	\end{itemize}
	Thus, the desired results are obtained and this completes the proof of Theorem \ref{theorem1}.
\end{proof}

\section{Proof of Lemma \ref{lemma6}}\label{GVR}
\begin{proof}
	On the communication round $r$, recall we have used $\frac{1}{N}\phi^t_i$ to estimate the global full gradient, i.e., $\phi^t_i=\sum_{n=1}^{N}\nabla f_n(\{z^t_n\})$. The update of $\phi^t_i$ only relies on the client $i$'s local dataset.  For simplicity and without loss of generalization, we set $S=1$, which means only one client is randomly participated in each round. Then, we have:
	\begin{equation}
		\mathbb{E}_i(\phi^t_i)=\frac{1}{N}\sum\nolimits_{i=1}^{N}\hat{\phi}^t_i+\frac{N-1}{N}\phi^r:=\frac{1}{N}\phi_t+\frac{N-1}{N}\phi^r.
	\end{equation}
	{ Moreover}, by straightforward calculation, it yields:
	\begin{equation}\label{proofofvariance}
		\begin{aligned}
			\mathbb{E}&\|{\tilde{g}^t_i-\mathbb{E}(\tilde{g}^t_i)}\|^2= \mathbb{E}\|{\phi^t_i-\frac{1}{N}\phi_t-\frac{N-1}{N}\phi^r}-g_{i,j_t}(z^t_{i,j_t})\\
			&\quad\quad\quad+\frac{1}{N}\sum\nolimits_{q=1}^{N}\nabla f_q(\{z^t_q\})+g_{i,j_t}(x^t_i)-\frac{1}{N}\sum\nolimits_{p=1}^{N}\nabla f_p(x^t_p)\|^2\\
			& \quad\quad\leq\frac{3(N-1)^2n_mL}{N}\cdot\sum\nolimits_{n,j}\mathbb{E}\| z^{i,t}_{n,j}-z^r_{n,j}\|^2+\frac{3n_mL}{N}\cdot\\
			&\quad\quad\quad\quad\quad\sum\nolimits_{n,j}\mathbb{E}\|z^{i,t}_{n,j}-x^t_{n}\|^2+3n^2_mL\mathbb{E}\|z^t_{n,j_t}-x^t_i\|^2,
		\end{aligned}
	\end{equation}
	where $n_m=\text{max}(\{n_i\}^N_{i=1})$, $z^{i,t}_{n,j}$ is trivially used to record the client $i$'s argument in $\phi^t_i$ at local iteration $t$, and $z^r_{n,j}$ is trivially utilized to record the arguments in $\phi^r$. Moreover, we have employed the relaxed triangle inequality in Lemma \ref{lemma1}  and assumed the $L$-smoothness of the $ f_{i,j}$ for the derivation. Since all the sequences $\{z^{i,t}_{n,j}\}$, $\{z^r_{n,j}\}$ and $\{x^t_i\}$ are from the aggregated global model $x^r, r=0,\dots,\infty$ during the update, hence when $x^r\rightarrow x^*$ in probability $1$, all the sequences will almost surely converge to $x^*$. This shows that the variance of the search direction $\tilde{g}^t_i$ in local update {progressively reduced to null}. 
\end{proof}

\section{FEDADMM FOR LOW RANK MATRIX ESTIMATION}\label{app:fedadmm}
In this section, we illustrate FedADMM for low rank matrix estimation. As shown in \cite{admm}, ADMM is a convenient and efficient tool for solving nonsmooth optimization problems. For the adaptation to FedOpt, we assume there are $N$ distributed clients, and denote $D_j\in\mathbb{R}^{d\times d}$ as the data sample.  Then  the local loss function of the low rank estimation on client $i$ is defined as: $f_i(X)=\sum_{j=1}^{n_i}(\langle X,D_j\rangle-y_j)^2$, where $y_j$ is the noisy observation. All clients target at solving the following problem via FedADMM:
\begin{equation}\label{lowrankestimation}
	\underset{X}{\text{min }} \sum_{i=1}^{N} f_i(X) + \lambda\norm{X}_*,
\end{equation}
Here, $f_i(X)$ has absorbed $\nicefrac{1}{N}$ for simplicity.  The problem  (\ref{lowrankestimation}) targets at recovering the low rank matrix $X\in\mathbb{R}^{d\times d}$ from the noisy observations $y_j$. For FedADMM, we first formulate its consensus form: 
\begin{equation}\label{consensus}
	\begin{aligned}
		&\underset{X}{\text{min }} \sum_{i=1}^{N} f_i(X_i) + \lambda\norm{Z}_*,\\
		&\text{s.t. } X_i-Z = 0, i=1,\dots,N,
	\end{aligned}
\end{equation}
where $Z\in\mathbb{R}^{d\times d}$ is the consensus variable, $X_i\in\mathbb{R}^{d\times d}$ can be viewed as the local model. Obviously, (\ref{consensus}) and (\ref{lowrankestimation}) are equivalent.  Then, the augmented Lagrangian function can be further derived \cite{admm}: 
\begin{equation}\label{lagrangian}
	 \mathcal{L}(X_{1:N}, \Pi, Z ) = \sum_{i=1}^{N} \left\{f_i(X_i) + \left\langle \pi_i, X_i-Z \right\rangle + \frac{\rho}{2}\norm{X_i-Z}^2_2\right\} + \lambda\norm{Z}_*,
\end{equation}
where $\pi_i\in\mathbb{R}^{d\times d}$ is the Lagrangian multiplier, $\left\langle , \right\rangle$ denotes the matrix inner product, $\rho>0$ is the regularization parameter and $\Pi=\{\pi_i,i=1,\dots,N\}$. Subsequently, classical ADMM \cite{admm} solves the problem (\ref{consensus}) via the following iterative procedure at the $t$th iteration:
\begin{equation}\label{Xi}
	X_i^{t+1}\leftarrow \underset{X_i}{\text{argmin}}f_i(X_i) + \left\langle \pi^t_i, X_i-Z^t \right\rangle + \frac{\rho}{2}\norm{X_i-Z^t}^2_2,\,\,\,\,\,\,\,\,\,\,\,\,\,\,\,\,\,\,\,\,\,\,\,\,\,\,\,
\end{equation}
\begin{equation}\label{pi}
	\pi_i^{t+1}\leftarrow \pi_i^{t}+\rho(X_i^{t+1}-Z^t),\,\,\,\,\,\,\,\,\,\,\,\,\,\,\,\,\,\,\,\,\,\,\,\,\,\,\,\,\,\,\,\,\,\,\,\,\,\,\,\,\,\,\,\,\,\,\,\,\,\,\,\,\,\,\,\,\,\,\,\,\,\,\,\,\,\,\,\,\,\,\,\,\,\,\,\,\,\,\,\,\,\,\,\,\,\,\,\,\,\,\,\,\,
\end{equation}
\begin{equation}\label{Z}
	Z^{t+1}\leftarrow \underset{Z}{\text{argmin}}\sum_{i=1}^{N} \left\{\left\langle \pi^{t+1}_i, X^{t+1}_i-Z \right\rangle + \frac{\rho}{2}\norm{X^{t+1}_i-Z}^2_2\right\} + \lambda\norm{Z}_*.
\end{equation}
Note that (\ref{Xi}) and (\ref{pi}) can be implemented in parallel over clients $i=1,\dots,N$. Moreover, for solving the subproblem (\ref{Xi}), we can use the efficient linearized approximation strategy for $f_i(X_i)$ at $X^t_t$, namely $f_i(X_i)\approx f_i(X_i^t)+\langle \nabla f_i(X^t_i),X_i-X^t_i\rangle+\nicefrac{1}{2\eta_l}\norm{X_i-X^t_i}^2_2$, where $\eta_l$ is the second-order approximate. Therefore,  the subproblem (\ref{Xi}) can be efficiently solved via:
\begin{equation}\label{X}
	X_i^{t+1}\leftarrow \frac{\rho\eta_lZ^t-\eta_l\pi^t_i+X^t_i-\eta_l\nabla f_i(X^t_i)}{1+\rho\eta_l}.
\end{equation}
Furthermore, for subproblem (\ref{Z}), it can be solved by using the proximal operator for the nuclear norm iteratively. To be specific, 
\begin{equation}\label{Proxz}
	Z^{m+1}\leftarrow  \text{ Prox}_{\lambda\eta_g\norm{\cdot}_*} \left\{Z^m-\eta_g\big(N\rho Z^{m}-\sum_{i=1}^{N}\{\pi^{t+1}_i+\rho X_i^{t+1}\}\big)\right\}
\end{equation}
is performed for $m=1,\dots,M$, then we can obtain $Z^{t+1}\leftarrow Z^M$.

As for FedADMM, it mimics the adaptation of SGD to FedAvg \cite{pmlr-v54-mcmahan17a}. To be specific, the participated clients $i\in\mathcal{S}$ performs the local updates (\ref{Xi}) and (\ref{pi}) for multiple times (say $T$ local iterations), and then the quantity $\pi_i^{r}+\rho X^{r}_i$ is transmitted to the server for performing the global update (\ref{Z}). While clients $i\in \mathcal{S}$ update their  local models $X_i$ and Lagrangian multipliers $\pi_i$, those clients $i\notin\mathcal{S}$ hold their $X_i$ and  $\pi_i$,  i.e., $X^{r}_i\leftarrow X^{r-1}_i$ and $\pi^{r}_i\leftarrow \pi^{r-1}_i$.  We summarize FedADMM in Algorithm \ref{alg:fedadmm}.
\begin{algorithm}[ht]
	\renewcommand{\algorithmicrequire}{\textbf{Input:}}
	\renewcommand{\algorithmicensure}{\textbf{Output:}}
	\caption{FedADMM} 
	\label{alg:fedadmm}
	\begin{algorithmic}[1]
		\STATE \textbf{server input:}  initial $Z^0$, $\eta_g$.
		\STATE \textbf{client $i$'s input:} initial $X^0_i$, $\pi_i^0$ and $\eta_l$.
		
		\FOR{$r=1,\dots,R$}
		\STATE 
		
		\underline{\textbf{Server implements} steps 5-6:}
		
		\STATE
		Obtain $Z^r$ by solving
		\begin{equation}\label{Zr}
			Z^{r}\leftarrow \underset{Z}{\text{argmin}}\sum_{i=1}^{N} \left\{\left\langle \pi^{r-1}_i, X^{r-1}_i-Z \right\rangle + \frac{\rho}{2}\norm{X^{r-1}_i-Z}^2_2\right\} + \lambda\norm{Z}_*.
		\end{equation}

		\STATE 
		
		Sample clients ${\mathcal{S}}\subseteq[N]$ and	transmit $Z^{r}$ to client $i\in{\mathcal{S}}$.
		
		\STATE
		\underline{\textbf{Clients implement} steps 8-14 \textbf{in parallel for} $i\in{\mathcal{S}}$:}
		
		\STATE
		After receiving $Z^r$, client $i\in{\mathcal{S}}$ performs
		
		\FOR{$t=1,\dots,T$}
		
		\STATE Sample index $j$ from $[n_i]$ for calculating the gradient $\tilde{\nabla} f_i(X^{t-1}_i)$.
		\STATE
		Obtain $X^r_i$ and $\pi^r_i$ respectively via:
		\begin{equation}\label{Xi1}
			X_i^{t}\leftarrow \frac{\rho\eta_lZ^r-\eta_l\pi^{t-1}_i+X^{t-1}_i-\eta_l\tilde{\nabla} f_i(X^{t-1}_i)}{1+\rho\eta_l}.
		\end{equation}
		\begin{equation}\label{pii}
			\pi_i^{t}\leftarrow \pi_i^{t-1}+\rho(X_i^{t}-Z^r),\,\,\,\,\,\,\,\,\,\,\,\,\,\,\,\,\,\,\,\,\,\,\,\,\,\,\,\,\,\,\,\,\,\,\,\,\,\,\,\,,
		\end{equation}
		\ENDFOR
		\STATE Set $X^r_i\leftarrow X^T_i$ and $\pi^r_i\leftarrow \pi^T_i$ for $i\in\mathcal{S}$, and $X^{r}_i\leftarrow X^{r-1}_i$ and $\pi^{r}_i\leftarrow \pi^{r-1}_i$ for $i\notin\mathcal{S}$.
		\STATE Client $i$ transmits $\pi_i^r+\rho X^r_i$ to the server.

		\ENDFOR
	\end{algorithmic}
\end{algorithm}

We have illustrated FedADMM and next, the experimental settings of FedADMM for low rank matrix estimation in Section \ref{sec:lrme} is described. To be specific, $\eta_l$ on each client is set to $\eta_l=10^{-4}$ and $\eta_g$ on the server is set to be $\eta_g=10^{-4}$. Moreover , the regularization parameter $\rho$ is tuned to $\rho=5$. In terms of solving (\ref{Zr}), $M=20$ proximal steps of (\ref{Proxz}) are performed. 

\end{document}